\documentclass[preprint,12pt]{elsarticle}
\usepackage{amssymb,amsmath}
\usepackage{hyperref}

\journal{Acta Astronautica}

\usepackage{color}

\begin{document}

\begin{frontmatter}

\title{The Deflector Selector: A Machine Learning Framework for Prioritizing Hazardous Object Deflection Technology Development}

\author[dtm,fdl]{E. R. Nesvold\corref{cor1}}
\author[ucla,fdl]{A. Greenberg}
\author[saao,fdl]{N. Erasmus}
\author[oxfd,fdl]{E. van Heerden}
\author[aten,mplc,fdl]{J. L. Galache}
\author[isc,fdl]{E. Dahlstrom}
\author[seti,fdl]{F. Marchis}

\cortext[cor1]{Corresponding author}
\address[dtm]{Department of Terrestrial Magnetism, Carnegie Institution for Science, 5241 Broad Branch Rd, Washington, DC 20015, USA}
\address[ucla]{Physics \& Astronomy Department, University of California, Los Angeles, 430 Portola Plaza, Box 951547, Los Angeles, CA 90095, USA}
\address[saao]{South African Astronomical Observatory, Cape Town, 7925, South Africa}
\address[oxfd]{Department of Physics, Oxford University, Parks Road, Oxford, OX1 3PU, UK}
\address[ftwt]{NASA FDL, 3248 W. 7th St, Apt 428, Fort Worth, TX 76107, USA}
\address[aten]{Aten Engineering, Portland, OR, USA}
\address[mplc]{Minor Planet Center, Harvard-Smithsonian Center for Astrophysics, Cambridge, MA, USA}
\address[isc]{International Space Consultants, 210 Waverly St \#6, Menlo Park, CA 94025, USA}
\address[seti]{SETI Institute, 189 Bernardo Ave, Mountain View, CA 94043, USA}
\address[fdl]{NASA Frontier Development Lab, Mountain View, CA, USA}

\begin{abstract}

Several technologies have been proposed for deflecting a hazardous Solar System object on a trajectory that would otherwise impact the Earth. The effectiveness of each technology depends on several characteristics of the given object, including its orbit and size. The distribution of these parameters in the likely population of Earth-impacting objects can thus determine which of the technologies are most likely to be useful in preventing a collision with the Earth. None of the proposed deflection technologies has been developed and fully tested in space. Developing every proposed technology is currently prohibitively expensive, so determining now which technologies are most likely to be effective would allow us to prioritize a subset of proposed deflection technologies for funding and development. We present a new model, the Deflector Selector, that takes as its input the characteristics of a hazardous object or population of such objects and predicts which technology would be able to perform a successful deflection. The model consists of a machine-learning algorithm trained on data produced by $N$-body integrations simulating the deflections. We describe the model and present the results of tests of the effectiveness of nuclear explosives, kinetic impactors, and gravity tractors on three simulated populations of hazardous objects.

\end{abstract}


\end{frontmatter}

\section{Introduction}
\label{sec:introduction}

Impacts on the Earth by natural Solar System objects (i.e., asteroids and comets) can pose a significant threat to human lives and infrastructure. Geological evidence of an impact crater near Chicxulub, Mexico indicates that an asteroid or comet 66 million years ago may have been responsible for the mass extinction of approximately 3/4 of the Earth's plant and animal species and the disruption of the global climate \cite{Schulte2010}. While the Chicxulub impactor was at least 10 km in diameter, impacting objects too small to cause mass extinctions can still pose a regional threat to life and property. Even the $\sim17$ m object that struck the atmosphere at a shallow angle over Chelyabinsk Oblast, Russia in 2013 caused \$33 million (USD) in infrastructure damage, and over $\sim$1,000 injuries \cite{Black2013}.

Several ground- and space-based observing campaigns, such as the Catalina Sky Survey \cite{Larson1998}, LINEAR (Lincoln Near-Earth Asteroid Program) \cite{Stokes2000}, Pan-STARRS (Panoramic Survey Telescope and Rapid Response System) \cite{Wainscoat2016}, NEOWISE (NEO Wide-field Infrared Survey Explorer) \cite{Mainzer2011} have been dedicated to detecting and tracking Near Earth Objects (NEOs) and flagging potentially hazardous NEOs for follow-up observations to characterize the probability that they will impact the Earth. The International Astronomical Union's Minor Planet Center\footnote{\url{http://www.minorplanetcenter.net/}}, the world's clearinghouse for asteroid and comet astrometry, provides coordination for the surveys searching for NEOS and maintains a database of asteroid and comet astrometric observations and orbits.

In the event that a hazardous object is detected on an Earth-crossing trajectory, several technologies have been proposed for deflecting the object to prevent an impact. Deflection technologies are designed to change the velocity of an impactor at some time before the impact, shifting the object's orbit such that the object and the Earth will not occupy the same position in space at the same time. Proposed deflection technologies include nuclear explosives \cite{Hammerling1995}, kinetic impactors \cite{Dachwald2007,Vasile2008}, gravity tractors \cite{Lu2005}, mass drivers \cite{Melosh1994}, laser ablation \cite{Lubin2016}, and ion beam shepherding \cite{Bombardelli2013}, among others. However, none of these technologies have been fully developed and tested in space. The kinetic impactor method has been indirectly attempted in space during the Deep Impact mission, which intentionally collided a 370 kg impactor with the comet Tempel 1, but the subsequent change in the velocity of the comet was not measured \cite{AHearn2005}. While nuclear explosives are a well-studied technology on Earth, the testing of nuclear explosives in space was prohibited by the Outer Space Treaty of 1967 \cite{Nations1966}. 

Theoretical studies of the various proposed deflection technologies have often focused either on modeling the capabilities of a single technology \cite{Hammerling1995,Lu2005}, or comparing the abilities of the different technologies to address specific impact scenarios \cite{Carusi2002,Vasile2008,SanchezCuartielles2010}. For example, Carusi et al. \cite{Carusi2002} performed $N$-body integrations of 11 objects on a collisional trajectory with the Earth and simulated the attempted deflection of the objects by applying a change in the object's velocity, $\Delta v$. They found that the $\Delta v$ necessary to deflect an object depends strongly on the lead time, the time before impact that the $\Delta v$ is applied. 

Developing and testing every proposed deflection technology would be prohibitively expensive. However, if humanity waits until a clear impact threat is detected to decide which technology to use, there may not be time to develop and deploy the chosen technology before impact. The readiness levels of proposed deflection technologies vary \cite{SanchezCuartielles2007}, but if the development time is longer than the lead time between the detection of a threat and its arrival, the technology will be useless. Even technologies with development times slightly longer than the lead time may not be deployed with enough time to be effective. Determining now which technologies are most likely to be useful would allow policy and funding decision-makers to effectively prioritize a subset of the proposed deflection technologies. 

We have developed a method for comparing the effectiveness of the various proposed technology on deflecting the likely population of hazardous objects. Our model, the Deflector Selector, maps the distribution of parameters of a hypothetical impactor population to the set of proposed technologies that can feasibly deflect these objects. In this work, we describe the Deflector Selector framework and use it to address the following questions:
\begin{itemize}
\item Which deflection method has the highest likelihood of deflecting the broadest range of potentially hazardous objects?
\item Which object characteristics is the choice of deflection method most sensitive to?
\item Which areas of the impactor parameter space are not covered by current deflection technologies?
\end{itemize}

Our model consists of a machine-learning algorithm that takes as its input the characteristics of a hypothetical hazardous object (e.g., orbital parameters, size, etc.) and outputs a list of the deflection technologies capable of deflecting the object. To train the algorithm, we produced a set of training data using orbital simulations to simulate the application of a change in velocity, $\Delta v$, to deflect a hazardous object. For each object, we attempt multiple deflections, each with a $\Delta v$ representing the maximum deflection achievable by one of the deflection technologies we consider. 

Section \ref{sec:orbital} describes our method for producing and characterizing a set of simulated hazardous objects. In Section \ref{sec:deflection}, we summarize the object deflection simulations we use to test the effectiveness of the three technologies we consider in this work. In Section \ref{sec:machinelearning}, we describe the machine learning algorithm we trained to predict which deflections technologies are most generally effective. We summarize our results in Section \ref{sec:results} and discuss our conclusions in Section \ref{sec:conclusions}.

\section{Preparing the Earth-Impacting Objects}
\label{sec:orbital}

\subsection{Simulating the Hazardous Object Population}
\label{sec:orbitpopulation}

The first step in simulating the effects of the various deflection technologies is to generate a population of hypothetical, undiscovered Earth-impacting objects. We characterize each hazardous object with five parameters: the $\beta$ parameter representing the object's internal structure, the object's diameter $D$, and the semi-major axis $a_{\rm obj}$, eccentricity $e_{\rm obj}$, and inclination $i_{\rm obj}$, representing the object's orbit. Table \ref{tab:object} summarizes these parameters and the range we used for each. 

\begin{table}[h]
\centering
\begin{tabular}{lc}
\hline
Parameter & Range \\ 
\hline
$a_{\rm obj}$ (au) 	& 0.5-2.0 	\\
$e_{\rm obj}$ 		& 0.0-1.0 	\\
$i_{\rm obj}$ (deg)	& 0-40 	\\
$D$ (m)	& 50-1000	\\
$\beta$	& 0-4	\\
\end{tabular}
\caption{The parameters we use for our simulated Earth-impacting objects, and their ranges. The semi-major axis $a_{\rm obj}$, eccentricity $e_{\rm obj}$, inclination $i_{\rm obj}$, and $\beta$ parameter are drawn from a uniform distribution, while the diameter $D$ follows a power-law distribution. Note that the eccentricity $e_{\rm obj}$ is limited for a given object's $a$ to ensure that the object's orbit is Earth-crossing.}
\label{tab:object}
\end{table}

The $\beta$ parameter is also known as the momentum enhancement factor. Values of $\beta$ greater than 1 indicate that when the object is struck by an impactor, the momentum of the object is increased due to the material expelled. Values less than 1 indicate that not all of the impactor's momentum is transferred to the object. We randomly draw the $\beta$ parameter values from a uniform distribution between $\beta = 0$ and $\beta = 4$. 

To select an object's diameter, we use a size distribution based on the Bottke et al. empirical absolute magnitude distribution \cite{Bottke2000}:
\begin{equation} n(H) \propto 10^{0.35 H} dH, \end{equation}
where $n(H)$ is the number of objects with absolute magnitudes between $H$ and $H+dH$. This absolute magnitude distribution can be converted into a size distribution using:
\begin{equation} \label{eqn:absmag} D = \frac{10^{-H/5}}{\sqrt{p}} \times 1329~{\rm km}, \end{equation}
where $D$ is the diameter of the object and $p$ is its albedo \cite{Fowler1992}. We assume an albedo of 0.25 for all objects and restrict our size range to objects with radii between 50 m and 1 km. We approximate the objects as spheres of constant density $3~{\rm g/cm^3}$ to estimate each object's mass. 

To produce an object's orbit, we select a semi-major axis $a_{\rm obj}$ and inclination $i_{\rm obj}$ from uniform distributions in the ranges $a_{\rm obj}=0.5-2$ au and $i_{\rm obj}=0-40^{\circ}$, respectively. We select an eccentricity $e_{\rm obj}$ from a uniform distribution, nominally in the range $e_{\rm obj}=0.0-1.0$. For a given semi-major axis value, this eccentricity range may be further restricted by the requirement that the object's orbit must be Earth-crossing, i.e., if $r_\oplus$ is the distance from the Earth to the Sun, we require $a_{\rm obj}(1-e_{\rm obj}) <= r_\oplus$ and $a_{\rm obj}(1+e_{\rm obj}) >= r_\oplus$. For each object, this set of $(a_{\rm obj},e_{\rm obj},i_{\rm obj})$ does not provide all of the information needed to describe the object's orbit. We must select a longitude of the ascending node $\Omega_{\rm obj}$, argument of perihelion $\omega_{\rm obj}$, and mean anomaly $M_{\rm obj}$ such that the object is co-located with the Earth at the time of impact, $t=0$, given the Earth's orbital elements $(a_\oplus, e_\oplus, i_\oplus, \Omega_\oplus, \omega_\oplus, M_\oplus)$. We perform this selection as follows.

The impact must occur where the object's orbit crosses the Earth's orbital plane. Using an ecliptic reference frame, this will occur at the object's ascending or descending node, so we first set the object's longitude of ascending node to the Earth's true anomaly, $\Omega_{\rm obj} = \nu_{\oplus}$. 

Our next step is to select a value for mean anomaly for the object, $M_{\rm obj}$, such that the object's distance to the Sun is equal to the Earth's, $r_{\rm obj} = r_{\oplus}$. We find this $M_{\rm obj}$ numerically by temporarily setting the object's argument of perihelion to zero $\omega_{\rm obj}=0$ and iterating through possible values of $M_{\rm obj}$ to minimize $\lvert r_{\oplus} - r_{\rm obj} \rvert$.

The final step is to find a value for $\omega_{\rm obj}$ such that $\vec{r_{\oplus}}$ and $\vec{r_{\rm obj}}$ are parallel, i.e., $\vec{r_{\oplus}} \cdot \vec{r_{\rm obj}} = \lvert \vec{r_{\oplus}} \rvert \lvert \vec{r_{\rm obj}} \rvert$. We find this $\omega_{\rm obj}$ numerically by iterating through possible values of $\omega_{\rm obj}$ numerically to minimize $\lvert \vec{r_{\oplus}} \cdot \vec{r_{\rm obj}} - \lvert \vec{r_{\oplus}} \rvert \lvert \vec{r_{\rm obj}} \rvert \rvert$. Our selection of $M_{\rm obj}$ requires that the object and the Earth are equidistant to the Sun, and our selection of $\omega_{\rm obj}$, requires that the position vectors of the Earth and the object are parallel. Together, these two orbital parameters ensure that the Earth and the object have the same position in space, i.e., are colliding.

To simulate an attempted deflection of the object, we need to know the orbital parameters of the object $(a'_{\rm obj}, e'_{\rm obj}, i'_{\rm obj}, \Omega'_{\rm obj}, \omega'_{\rm obj}, M'_{\rm obj})$ and the Earth $(a'_{\oplus}, e'_{\oplus}, i'_{\oplus}, \Omega'_{\oplus}, \omega'_{\oplus}, M'_{\oplus})$ at time $t=-t_{\rm lead}$ before the impact at time $t=0$. In a purely Keplerian system, only the mean anomalies would change with time:
\begin{equation} M'_{\rm obj} = M_{\rm obj} - \frac{t_{\rm lead}\sqrt{GM_{\odot}}}{2\pi a_{\rm obj}^{2/3}}, \end{equation}
\begin{equation} M'_{\oplus} = M_{\oplus} - \frac{t_{\rm lead}\sqrt{GM_{\odot}}}{2\pi a_{\oplus}^{2/3}}, \end{equation}
where $G$ is the gravitational constant and $M_{\odot}$ is the mass of the Sun. However, gravitational perturbations from the Earth and other planets in the simulation will change all of the object's orbital parameters with time and must be numerically calculated. Therefore, instead, we integrate the orbits of the object backwards in time using a leapfrog $N$-body integrator (see Section \ref{sec:nbody}). Leapfrog integrators are symmetric in time, guaranteeing that the object will impact the Earth at time $t=0$ in the forward integration. In the simulations discussed in this work, we integrated backwards in time to $t=-15$ yr before impact.

\subsection{Backwards Integration}
\label{sec:nbody}

We performed orbital simulations using the $N$-body integrator REBOUND \cite{Rein2012}, run on the Carnegie Institute of Washington's Memex cluster. Our simulations included the gravitational effects of Jupiter, Venus, Mars, and Earth's Moon, as well as the Sun and the Earth. We used the included leapfrog integrator in REBOUND with a constant timestep of $\Delta t=1.6\times10^{-6}$ yr, and a gravitational softening parameter of $r_s=R_{\oplus}$.

As described in Section \ref{sec:orbitpopulation}, determining the initial orbital parameters of an object at time $t=-t_{\rm lead}$ that guarantee a collision with the Earth at $t=0$ is a non-analytic problem in the presence of gravitational perturbations from the Earth and other planets. We address this problem numerically by first placing the object at the location of the Earth at time $t=0$ and then integrating the system backwards for 15 yr. The resulting system coordinates are the appropriate initial conditions for the forward simulation to guarantee a collision at $t=0$, due to the time-reversibility of the leapfrog integrator.

To simulate an attempted deflection of an Earth-impacting object, we will apply a change in velocity, $\Delta v$, to the object in the direction of the object's instantaneous velocity at some lead time $t=-t_{\rm lead}$. We describe our calculation $\Delta v$ in detail in Section \ref{sec:deflection}. The lead time $t_{\rm lead}$ represents the time between the application of the deflection technology and the Earth-impacting time. The object's orbit places an upper limit on the lead time in two ways. First, the object must be detected and its orbit measured before the deflection technology can be deployed. Second, the technology must travel from the Earth to an intercept point with the object in order to attempt a deflection. During the backwards integration, we simulate the detection of each hazardous object and calculate the time needed for the deflection technology to reach the object, to explore the dependence of the lead time on the object's orbit and size. We describe the detection simulation and travel time simulation in the following subsections.

\subsubsection{Detection Simulation}
\label{sec:detection}

We simulate the detection of the object by defining a ``detection zone'' traveling with the Earth through space. During the backwards integration, we record the times at which the object crosses into and out of this detection zone. The boundaries of our simulated deflection zone for a given object are set by the magnitude limit and sky coverage of the detector. The regions of the sky covered by a ground- or space-based instrument vary widely, so for now let us consider a hypothetical detector located at the Earth that can detect any object with a magnitude brighter than $m_{\rm max}$ but cannot point at the Sun, i.e., it cannot detect any object within $\theta_{\rm FOV}$ of the Sun, where $\theta_{\rm FOV}$ is the field of view of the detector.

At every timestep of the backwards integration, we estimate the object's apparent magnitude as measured from Earth based on its absolute magnitude $H_{\rm}$:
\begin{equation} m_{\rm obj} = H + 5\log(r_{\rm obj} r_{\rm dist}) - 2.5 \log(\phi(\theta)), \end{equation}
where $r_{\rm obj}$ is the distance between the object and the Sun, $r_{\rm dist}$ is the distance between the object and the Earth, $\theta$ is the angle formed by the object-Sun line and the object-Earth line, and $\phi$ is the object's phase function. We use a phase function for an ideal diffuse reflecting sphere,
\begin{equation} \phi(\theta) = \frac{2}{3} \left( \left( 1-\frac{\theta}{\pi} \right) \cos \theta + \frac{1}{\pi} \sin \theta \right). \end{equation}
We define an object to be in the detection zone if $m_{\rm obj} < m_{\rm max}$ and $\theta_{\oplus} > \theta_{\rm FOV}$, where $\theta_{\oplus}$ is the angle formed by the Earth-Sun line and the object-Earth line. We set our minimum elongation angle to $\theta_{\rm FOV} = 60^{\circ}$ and our apparent magnitude limit to $m_{\max} = 20.5$ \cite{Galache2015}. As a zeroth-order approximation, we consider the earliest time in which the object is in the detection zone to be the time at which the object is discovered and its orbit measured. In reality, multiple observations are required to measure the orbit of a hazardous object and confirm that it will impact the Earth.  This discovery time, $t = -t_{\rm disc}$, places an upper limit on the possible lead time: $t_{\rm disc} > t_{\rm lead}$. 

\begin{figure}
\includegraphics[width=\linewidth]{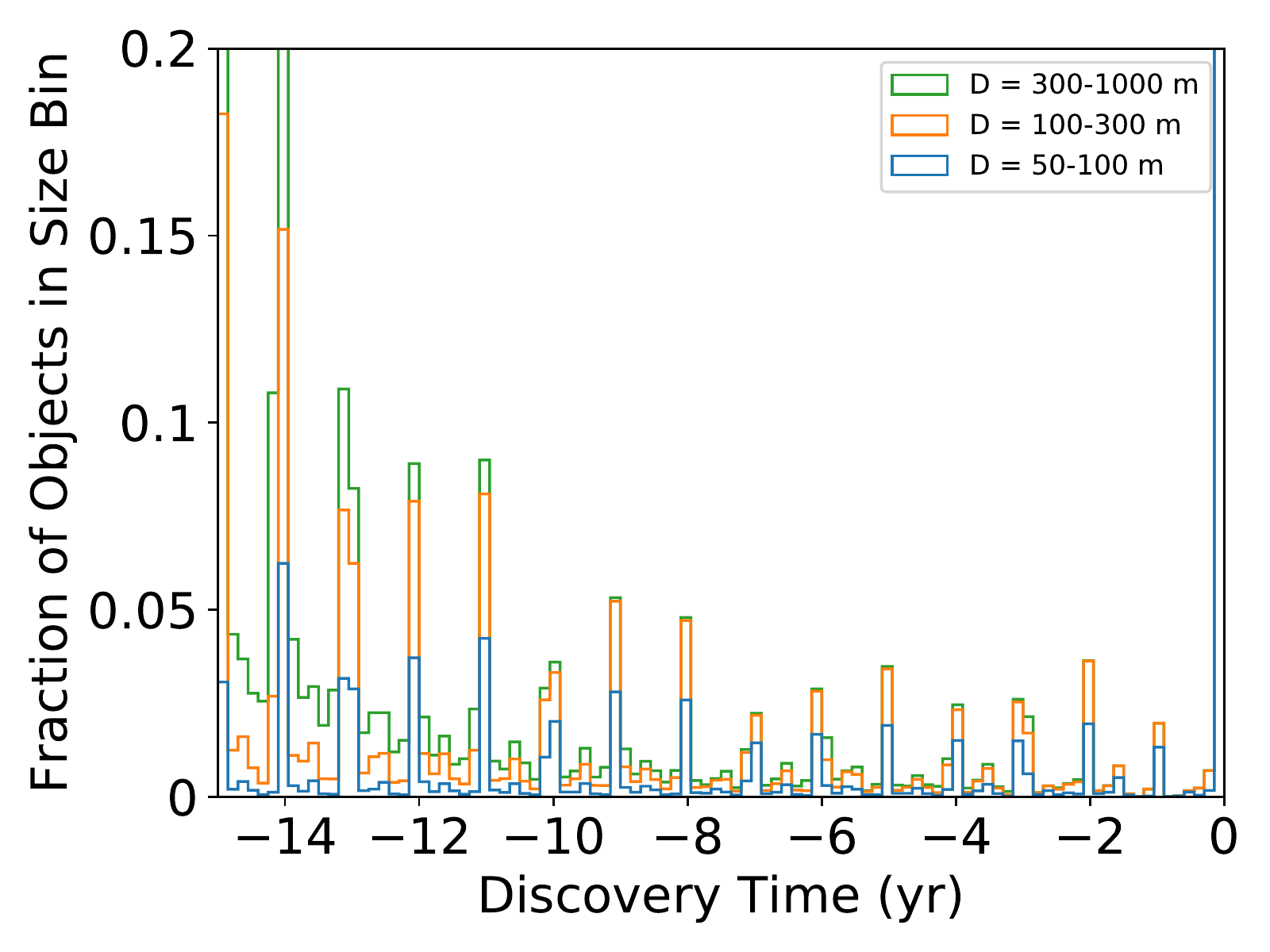}
\caption{\label{fig:disctimehist} Histograms of the discovery time of objects in our simulations, sorted into three diameter bins. The peaks represent close passes of the objects in the years before impact, during which detection is more likely. Larger objects are more likely to be discovered earlier.}
\end{figure}

The discovery time depends on both the object's orbit and its size. Figure \ref{fig:disctimehist} illustrates both of these effects, in a histogram of the discovery time of the objects in our simulations. The peak at $t_{\rm disc} = 15$ yr represents the significant fraction of all objects discovered at the earliest possible time, most of which would likely be discovered earlier if our simulations covered a larger range in time. Another peak close to time $t = 0$ yr represents the objects that are not discovered until their final approach to the Earth in the last year before impact. The remaining oscillation in the discovery time curves reflects the clustering of the mean anomalies of the objects. We have created a population of objects that all impact the Earth at time $t=0$. The peaks in the discovery time curves occur during close flybys of the objects in the years before detection. The histograms in Figure \ref{fig:disctimehist} are sorted into three object diameter bins. Larger objects are more likely to be discovered earlier before impact, as shown by the larger amplitude of the peaks in the discovery time curve.

\subsubsection{Travel Simulation}
\label{sec:travel}

The second constraint on lead time is the finite time needed for the technology to travel from the Earth to an intercept point with the hazardous object. If we assume that our simulated human civilization responds instantly to the discovery of an impact threat by launching the deflection technology, then this travel time $t_{\rm trav}$ can be used to further constrain the lead time: $t_{\rm disc} - t_{\rm trav} \geq t_{\rm lead}$.

We assume that our simulated human civilization will attempt to deflect the object at a point where the object crosses the Earth's orbital plane (i.e., the object's ascending or descending node), assuming enough time remains for the deployment technology to intercept the object at one of its nodes. While previous studies have shown that applying a $\Delta v$ at an object's perihelion is most effective for deflecting the object \cite{Ahrens1992}, the dynamical cost of reaching perihelion may make this choice of intercept location less effective \cite{Hall1997}. For example, more eccentric objects will have perihelion locations closer to the Sun, increasing the launch velocity required to reach that point, and potentially preventing a rocket from reaching perihelion within the lead time available. Instead, we select a point in Earth's orbital plane, reducing the energy required for the rocket to intercept the object and increasing the capacity of the rocket.

Suppose the locations of the Earth and the incoming hazardous object at the time of the object's discovery, $t=-t_{\rm disc}$, are $\vec{r}_{\rm Earth}(-t_{\rm disc})$ and $\vec{r}_{\rm imp}(-t_{\rm disc})$, respectively. If the time required for the object to reach its next node (ascending or descending) is $t_{\rm node}$, then the impactor will pass through its node before it impacts the Earth if $t_{\rm disc}-t_{\rm node}\geq 0$. The object's location at this node can be written as $\vec{r}_{\rm imp}(-t_{\rm disc}+t_{\rm node})$.

We are then confronted with Lambert's problem: we require a transfer orbit that can take the deflection technology from the Earth's location at time of discovery, $\vec{r}_{\rm Earth}(-t_{\rm disc})$ to the object's location as it passes through its node, $\vec{r}_{\rm imp}(-t_{\rm disc}+t_{\rm node})$ in the time it takes the object to reach the node, $t_{\rm node}$. More formally, we need to find a solution for the differential equation
\begin{equation} \frac{d^2\vec{r}}{dt^2} = -\frac{\mu}{r^3} \vec{r} \end{equation}
such that $\vec{r}(t_0) = \vec{r}_{\rm Earth}(t_0)$ and $\vec{r}(t_1) = \vec{r}_{\rm imp}(t_1)$, where $t_0 = -t_{\rm disc}$ and $t_1 = -t_{\rm disc}+t_{\rm node}$.

Solving Lambert's equation reveals the velocity change, $v_{\rm launch}$, necessary to place the deflection technology on this transfer orbit. To solve Lambert's equation, we use the universal variable method derived in Bate, Mueller, and White \cite{Bate1971} and Vallado \cite{Vallado1997}. There are practical limits on the launch velocity $v_{\rm launch}$ of a rocket, based on the mass of the payload and the rocket's capabilities. We discuss these limits in detail in Section \ref{sec:launch}. For now, we place an upper limit of $v_{\rm rocket} = 16$ km/s, approximately the Solar System escape velocity relative to the Earth and the speed at which the New Horizons spacecraft was launched \cite{NewHorizons2007}.

For every incoming hazardous object and corresponding discovery time $t_{\rm disc}$ produced via our backwards integration (described in Section \ref{sec:detection}), we use the following algorithm to calculate the travel time necessary for the deflection technology to reach the object:

\begin{enumerate}
\item Calculate $t_{\rm node}$, the time required for the object to travel from its location at discovery, $\vec{r}_{\rm imp}(-t_{\rm disc})$, to its next ascending or descending node, $\vec{r}_{\rm imp}(-t_{\rm disc}+t_{\rm node})$.
\item \label{step:perihelion} If $t_{\rm disc}-t_{\rm node}\geq 0$, then the object will pass its next node before it impacts the Earth, so enough time remains for humanity to attempt to intercept the object at a node crossing. Set the travel time to $t_{\rm trav} = t_{\rm node}$, and proceed to Step \ref{step:lambert}. Otherwise, proceed to Step \ref{step:toolate}.
\item \label{step:lambert} Using the chosen travel time $t_{\rm trav}$, solve Lambert's equation to calculate the launch velocity, $v_{\rm launch}$, necessary to place deflection technology onto transfer orbit. 
\item \label{step:goodrocket} Compare required launch velocity with the maximum possible launch velocity achievable with current technology, $v_{\rm rocket}$. If $v_{\rm launch}\leq v_{\rm rocket}$, then this transfer orbit is achievable. Record travel time and launch velocity. Otherwise, proceed to Step \ref{step:badrocket}.
\item \label{step:badrocket} If $v_{\rm launch} > v_{\rm rocket}$, then the deflection technology cannot intercept the object at its next node crossing due to limitations in launch capabilities. Calculate the time at which the object will cross the Earth's orbital plane once more. Using this new time as $t_{\rm node}$, return to Step \ref{step:perihelion}. If no node crossings remain before impact with the Earth, proceed to Step \ref{step:toolate}.
\item \label{step:toolate} If $t_{\rm disc}-t_{\rm node}<0$, then the object will not pass through one of its ascending or descending nodes before impacting the Earth. Alternatively, if the object will pass a node before impact but the required launch speed to reach the node is not achievable, then the deflection technology cannot intercept the object at its ascending or descending node. In either case, our simulated human civilization will then attempt to intercept the object as soon as possible before impact. We first set the travel time to half the remaining time until impact, $t_{\rm trav} = t_{\rm disc}/2$ or, in the case where we have already attempted to intercept at a node, the time between the final node crossing and the time of impact. We then repeat this algorithm from Step \ref{step:lambert}, resetting the travel time in each iteration to equal half the remaining time until impact, until either the inequality in Step \ref{step:goodrocket} is satisfied or ten iterations have passed. In this latter case, our simulated human civilization is incapable of reaching the incoming hazardous object with a deflection technology before impact, and we record the object as a failed deflection.
\end{enumerate}

This algorithm will result in either a failed deflection, which we record as such without the need for further simulation, or a travel time $t_{\rm trav}$ in which the deflection technology can reach the incoming hazardous object before impact, and a launch velocity $v_{\rm launch}$ necessary to place the technology on its transfer orbit. We record $v_{\rm launch}$ and set the lead time for the object as $t_{\rm lead}=-t_{\rm disc}+t_{\rm trav}$. 

\begin{figure}
\includegraphics[width=\linewidth]{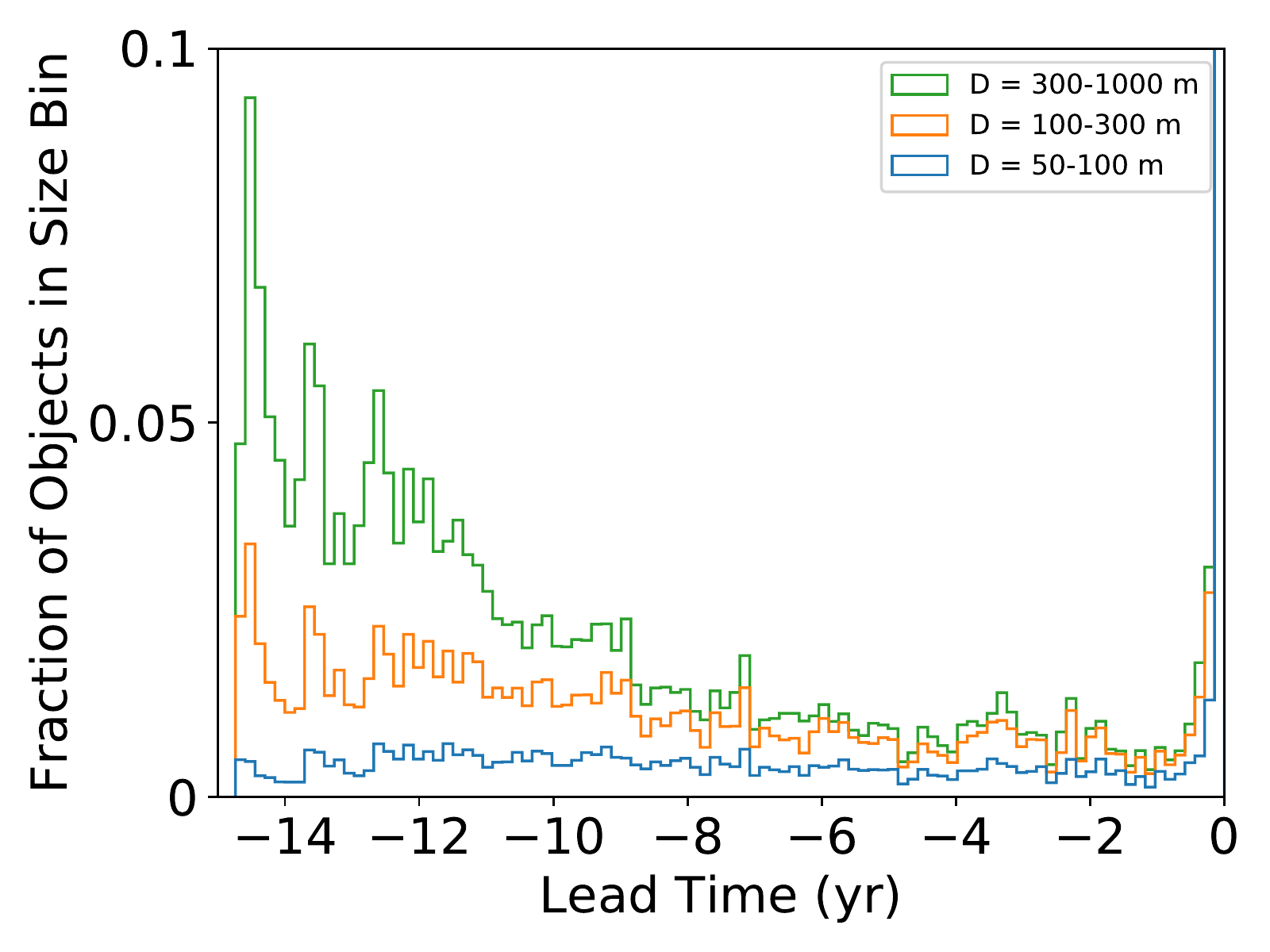}
\caption{\label{fig:leadtimehist} Histograms of the lead time of all objects in our simulations, sorted into three size bins. The peaks seen in the discovery time histograms in Figure \ref{fig:disctimehist} are smoothed somewhat by the addition of the travel time, which has a complex dependence on the orbital configuration of each object. The lead time for larger objects tends to be earlier, since they tend to be detected earlier.}
\end{figure}

Figure \ref{fig:leadtimehist} shows the lead time histograms for the objects in our simulations, each on its own orbit within the range specified by Table \ref{tab:object}. The peaks seen in the discovery time curves are smoothed out somewhat in the lead time curves due the the addition of the travel time, which depends on the orbit and position of the object relative to the Earth in a complex way. However, the effect of the object's size remains: larger objects, which tend to be detected earlier, can therefore be reached sooner with deflection technologies.

\begin{figure}
\includegraphics[width=\linewidth]{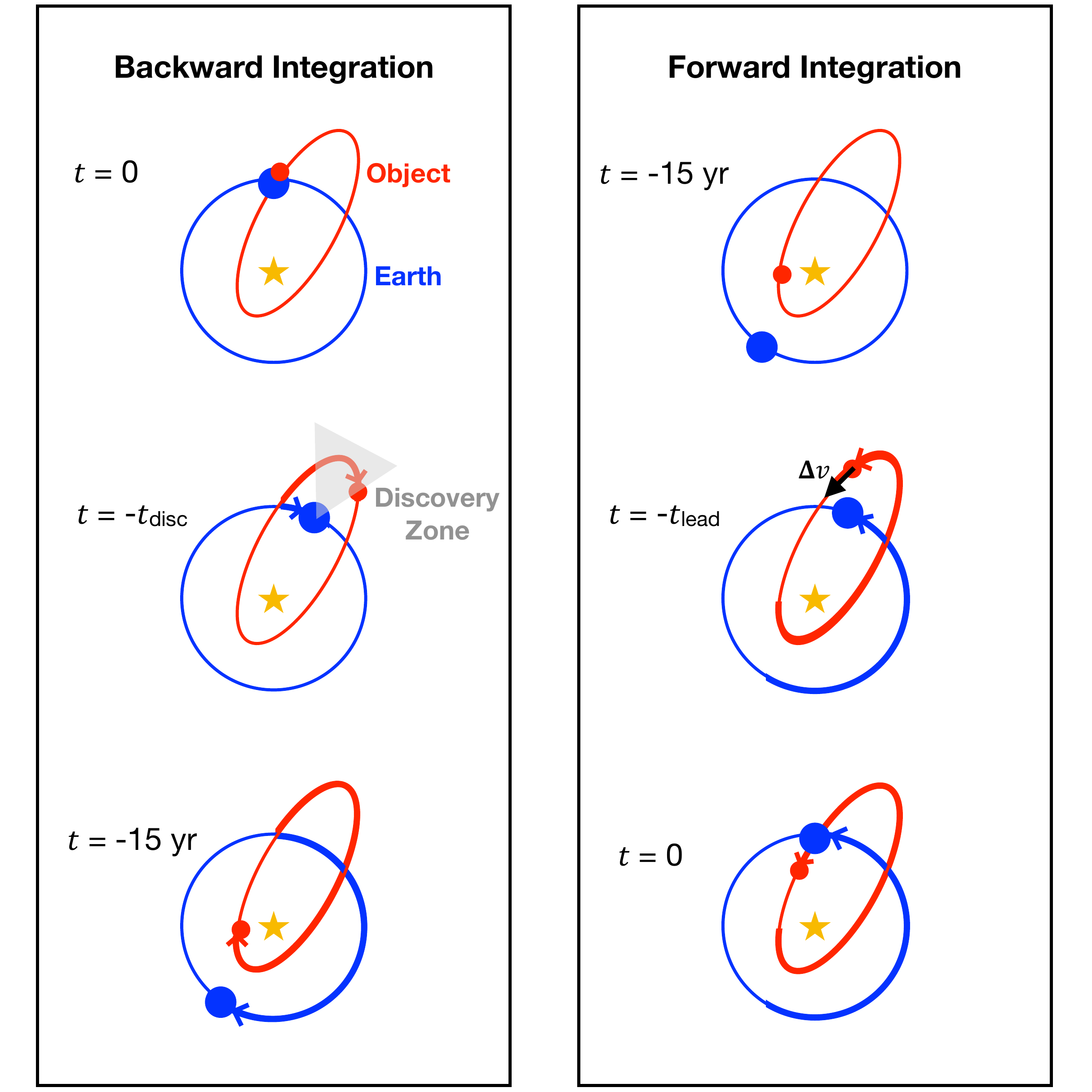}
\caption{\label{fig:diagram1} A diagram illustrating the integration steps of the Deflector Selector algorithm (orbital periods and sizes not to scale). The hazardous object is given an orbit such that it is colliding with the Earth at time $t=0$. The orbits of the object and the planets are then integrated backwards for 15 yr. The discovery time $t_{\rm disc}$ is recorded at the earliest time in which the object is discoverable from the Earth. Next, the object and planets are integrated forward in time from $t=-15$ yr. At some lead time $t = -t_{\rm lead}$, a $\Delta v$ is applied to the object to simulate a deflection attempt. If the object does not impact the Earth during the forward integration, the simulation is recorded as a successful deflection.}
\end{figure}

Figure \ref{fig:diagram1} illustrates the backward integration steps described in this section. At the end of the backwards integration, the object's orbital parameters are recorded. The lead time $t_{\rm lead}$ is calculated from the discovery time, $t_{\rm disc}$, found during the backwards integration, and the travel time required, $t_{\rm trav}$, estimated with the method described above. In the following section, we describe the next steps in the simulation: integrating the orbits forward, simulating an attempted deflection by applying a $\Delta v$ to the object, and checking for impacts.

\section{Simulating the Deflections}
\label{sec:deflection}

For each of our simulated hazardous objects (described in Section \ref{sec:orbital}), we can use the required launch velocity $v_{\rm launch}$ to estimate the maximum mass of the deflector that our simulated human civilization could launch to the incoming hazardous object. This upper limit on mass constrains the $\Delta v$ that a given deflection technology can apply. After calculating these $\Delta v$ values, we perform $N$-body simulations (with time running forward) to simulate the application of each $\Delta v$ (see Figure \ref{fig:diagram1}). Section \ref{sec:launch} below describes our calculations of the maximum deflector mass. Section \ref{sec:technologies} summarizes the capabilities of the three deflection technologies we consider in this work: a nuclear explosive, a kinetic impactor, and a gravity tractor. Section \ref{sec:deflectionsim} describes the forward integration that simulates a deflection attempt and tests for its effectiveness.

\subsection{Launch Vehicles}
\label{sec:launch}

The successful execution of any chosen deflection method is ultimately dependent on launching the technology into space and into the necessary transfer orbit. We assume that only one technology is used for an attempted deflection of an incoming hazardous object, and that only a single launch vehicle is used for the deployment of that technology. Our travel time estimation, described in Section \ref{sec:travel}, uses Lambert's equation to predict the velocity relative to the Earth, $v_{\rm launch}$, necessary to launch the deflection technology into the transfer orbit. We can then use the Tsiolkovsky ideal rocket equation to estimate the mass ratio of the launch vehicle: 
\begin{equation} \label{eqn:rocket} v_{\rm launch} = I_{\rm sp} g \ln \frac{m_{\rm init}}{m_{\rm final}}, \end{equation}
where $I_{\rm sp}$ is the specific impulse of the rocket, $g=9.8~{\rm m}/{\rm s}^2$ is the acceleration due to gravity on Earth, $m_{\rm init}$ is the total initial mass of the rocket before launch, $m_{\rm final}$ is the mass of the rocket after the propellant has been expended, also known as the dry mass, and we have assumed that the launch vehicle is a single-stage rocket. The total initial mass of the rocket is made up of the propellant, the payload, and the vehicle itself. We consider the mass of the vehicle to be part of the payload mass in this context, since two of our technologies (the nuclear explosive and the kinetic impactor) use the spacecraft itself to impart momentum on the target object via a collision, and the third (the gravity tractor) uses the total mass of the spacecraft to exert a gravitational pull on the target object. We therefore use an initial mass of $m_{\rm init} = m_{\rm propellant} + m_{\rm payload}$ and a final mass of $m_{\rm final} = m_{\rm payload}$. We roughly approximate the characteristics of the Delta IV Heavy, currently the orbital launch system with the largest payload capacity, setting $m_{\rm propellant} = 7\times10^{5}$ kg and $I_{\rm sp} = 420$ s. 
Rearranging Equation \ref{eqn:rocket} and inserting these values, we estimate that the maximum payload mass that can be launched with relative velocity $v_{\rm launch}$ is
\begin{equation} \label{eqn:payload} m_{\rm payload} = \frac{7\times10^5~{\rm kg}}{e^{v_{\rm launch}/(4116~{\rm m/s})} -1}, \end{equation}
For each deflection simulation produced by Section \ref{sec:orbital}, we use Equation \ref{eqn:payload} to estimate the maximum mass of the deflection technology. 

\subsection{Deflection Technologies}
\label{sec:technologies}

In this work, we selected three deflection technologies to study: the explosion of a nuclear device on or near the object, a collision with a kinetic impactor, or the alteration of the object's trajectory with a gravity tractor. We consider these to be three of the more plausible of the technologies discussed in Section \ref{sec:introduction}. Nuclear explosive technology was developed decades ago and has been tested and deployed on Earth, although testing in space was halted by the Outer Space Treaty of 1967 \cite{Nations1966}. Kinetic impactors are one of the most low-tech solutions proposed, requiring only the ability to rendezvous with a hazardous orbit, a technique which is required for all of the proposed technologies and which was demonstrated in space by the Deep Impact mission \cite{AHearn2005}. The gravity tractor, while requiring more complex technologies for the deployment and stationkeeping of the spacecraft, also relies on well-known physics. Deflection methods that require landing on the object's surface, such as a mass driver, pose a more significant technological challenge and depend strongly on the rotation of the object, which is difficult to measure.

Deflection via laser ablation would also depend strongly on the object's rotation, and would require the development of a powerful enough laser on a spacecraft with precise stationkeeping and aim. The remaining deflection technology discussed in Section \ref{sec:introduction}, ion beam shepherding, is similar to the gravity tractor in that it is a slow-push technique that depends only upon existing technologies. However, due to constraints in computation time, we chose to consider only the gravity tractor as a representative slow-push technology.

For each of these three technologies (nuclear explosive, kinetic impactor, and gravity tractor), we performed a literature search, summarized below, to investigate how the $\Delta v$ that each technology can apply depends on the object's size $D$, the object's $\beta$ parameter, and the mass of the deflector $m_{\rm payload}$.

\subsubsection{Nuclear Explosive}
\label{sec:nuclear}

Although nuclear explosives are highly controversial and subject to international treaties \cite{Nations1966}, they remain an extremely attractive option for deflection due to the high ratio of $\Delta v$ imparted to the target object, compared to other techniques. Nuclear explosives also stem from well-established and existing technology. On the other hand, the possibility of a catastrophic malfunction during launch or atmospheric travel poses a significant risk to the population surrounding the launch site and beyond. Until these issues are addressed, this method will remain a divisive topic. 

To harness the energy of a nuclear detonation to impart a $\Delta v$ on a target object, there are three possible approaches: a stand-off, surface, and sub-surface detonation. For this study we focus on the stand-off detonation method because it has a lower execution complexity and therefore a more realistic implementation, but should still be representative of the deflecting power of a nuclear explosion.

Nuclear radiation as a possible deflection method for NEOs has been extensively modeled \cite{Hammerling1995,SanchezCuartielles2010}. The velocity change imparted on the target object is the sum of the $\Delta v$ imparted by the x-rays, neutrons, and gamma rays produced by the nuclear explosion, and the smaller contribution of the debris from the payload capsule colliding with the target object. The total $\Delta v$ expected from a stand-off nuclear detonation is then
\begin{equation}
\Delta v_{\rm nuc} = \Delta v_{\rm x-ray} +  \Delta v_{\rm neutron} +  \Delta v_{\rm gamma} +  \Delta v_{\rm deb}.
\end{equation} 
The $\Delta v$ associated with each type of radiation is given by 
\begin{equation}
\Delta v_{\rm rad} = \frac{2\sqrt{2}}{\frac{1}{3} \rho r_{\rm obj} \mu_{\rm rad}} \sqrt{\epsilon_v} \left( \sqrt{F^*_{0,\rm rad}-1}-\arctan(\sqrt{F^*_{0,\rm rad}-1}\right),
\end{equation}
where the ``rad'' subscript indicates x-ray, neutron, or gamma, $\frac{1}{3} \rho r_{\rm obj}$ represents the area mass density of the target object (given the object's bulk density $\rho$, radius $r_{\rm obj}$, and approximating the object as a sphere), $\mu_{\rm rad}$ is the mass absorption coefficient for the given radiation type, and $\epsilon_v$ is the vaporization energy per unit mass \cite{Hammerling1995}. We use the vaporization energy per unit mass of $\epsilon_v = 8\times10^6$ J/kg adopted by Hammerling \& Remo \cite{Hammerling1995} (see Table \ref{tab:nuke} for a listing of the other parameters we used for each contributor to total $\Delta v$). $F^{*}_{0,\rm rad}$ is a unitless value associated with the given radiation type, given by
\begin{equation}
F^{*}_{0,\rm rad} = \frac{\eta_{\rm rad}E\Delta\Omega}{4\pi r_{\rm obj}^2} \frac{\mu_{\rm rad}}{\epsilon_v},
\end{equation}
where $\eta_{\rm rad}$ is the fraction of the total yield contributing the given radiation type, $E$ is the total energy yield of the nuclear explosion, and $\Delta\Omega$ is the fractional solid angle of radiation that impedes onto the asteroid \cite{Hammerling1995}. The latter is given by 
\begin{equation} \label{eqn:solidangle} \Delta\Omega = \frac{1}{2} - \frac{\sqrt{H}}{2} \frac{\sqrt{H+2r_{\rm obj}}}{H +r_{\rm obj}}, \end{equation}
where $H$ is the distance between the explosive and the target object at the time of detonation \cite{SanchezCuartielles2010}. Hammerling \& Remo \cite{Hammerling1995} found that the optimal stand-off distance in their model is $H=(\sqrt{2}-1)r_{\rm obj}$. While later work determined that the optimal stand-off distance has a more complex dependence on excess velocity and mass expelled \cite{SanchezCuartielles2010}, we use the Hammerling \& Remo expression for stand-off distance for simplicity.

Although in general, only a small contribution on the momentum change of the target object can be expected from the impact of the payload capsule debris, for completeness this velocity change $\Delta v_{\rm deb}$ can be expressed by
\begin{equation} \label{eqn:nuclear}
\Delta v_{\rm deb} = \frac{\beta S_{sc} m_{\rm deb} v_{\rm deb}}{m_{\rm obj}},
\end{equation}
where $\beta$ is a parameter of the target object describing how efficiently momentum is transferred to the object, which we conservatively set to $\beta=2$ \cite{SanchezCuartielles2010}, $S_{sc}=2/\pi$ is the scattering angle of the debris, $m_{\rm obj}$ is the mass of the target object, and $m_{\rm deb}$ and $v_{\rm deb}$ are the mass and velocity of the debris, respectively, which can be calculated as follows:
\begin{equation}
m_{\rm deb} = \Delta\Omega m_{\rm payload}
\end{equation}
\begin{equation}
v_{\rm deb} = \sqrt{\frac{2\eta_{\rm deb}E}{m_{\rm payload}}},
\end{equation}
where $\Delta\Omega$ is the fraction of the debris that hits the target object and is given by Equation \ref{eqn:solidangle} above, $m_{\rm payload}$ is the mass of the payload capsule, and $\eta_{\rm deb}$ is the fraction of the total nuclear yield contributing to the debris acceleration \cite{SanchezCuartielles2010}. 

\begin{table}[h]
\centering
\begin{tabular}{lcccc}
\hline
 & X-Ray & Neutron & Gamma Ray & Debris \\ 
\hline
$\eta$					&  0.7	& 0.01 				& 0.02			 	& 0.2 \\
$\mu~({\rm m}^2/{\rm kg})$	& 1.37	& $4.96\times10^{-3}$	& $2.34\times10^{-3}$ 	& -- \\ 
\end{tabular}
\caption{Parameters for our nuclear explosive calculations. Values for the contribution fraction $\eta$ assume a fission device \cite{Hammerling1995}. Values for the mass absorption coefficient assume a forsterite composition for the target object \cite{SanchezCuartielles2010}.}
\label{tab:nuke}
\end{table}

\begin{figure}
\includegraphics[width=\linewidth]{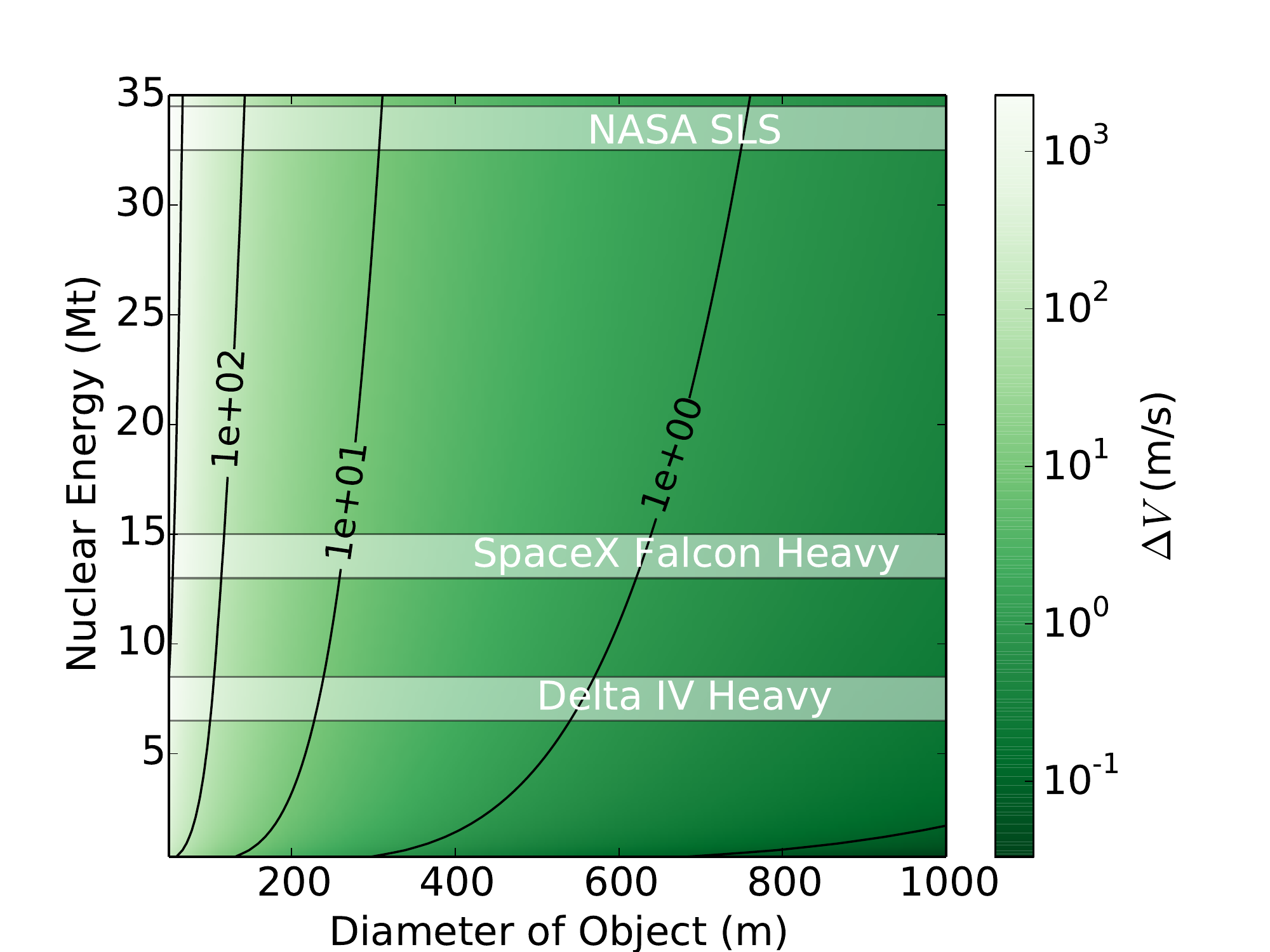}
\caption{\label{fig:deltavnuclear} Predicted $\Delta v$ imparted onto a target object by a stand-off nuclear explosion vs. object size and nuclear energy. Black solid lines trace contours in $\Delta v$. The nuclear energies that can be deployed by three current or proposed rockets are labeled, assuming a payload of B83 warheads delivered to a Mars orbit.}
\end{figure}

As an illustration, Figure \ref{fig:deltavnuclear} shows the predicted $\Delta v_{\rm nuc}$ values for a stand-off nuclear detonation as a function of the size of the target object and the potential nuclear energy. The range of nuclear energy is chosen to reflect the maximum payload of nuclear explosives that can be launched by various current and proposed heavy-lift rockets. For this rough calculation, we assume that a payload contains B83 warheads, each weighing 1.1 tons and capable of delivering 1.2 Mt of nuclear energy, and that the payload is delivered to a Mars orbit. The expected $\Delta v_{\rm nuc}$ values for the given range of target objects and nuclear energies range from $\sim10$ cm/s to a few km/s. 

\subsubsection{Kinetic Impactor}
\label{sec:kinetic}

A kinetic impactor imparts a velocity change in the target object simply by colliding a massive payload with the object to change the object's momentum. The appeal of this technique is that it is the least complex deflection approach. However, deflection via kinetic impact has not been tested in space and the details of the mechanism are poorly constrained. The $\beta$ parameter of the target object (previously described in Section \ref{sec:nuclear}), which measures how efficiently the momentum of the kinetic impactor is transferred to the object, is difficult to predict based on observations. This parameter can be highly dependent on both the internal structure and composition of the object. There is also a possibility that some of the kinetic energy of the impactor will fracture the target object into smaller (but still hazardous) fragments rather than fully deflecting the object's trajectory.

The $\Delta v_{\rm kin}$ imparted by a kinetic impactor can be estimated as follows:
\begin{equation} \label{eqn:kinetic}
\Delta v_{\rm kin} = \beta \frac{m_{\rm payload}}{m_{\rm payload}+m_{\rm obj}}v_{\rm rel},
\end{equation}
where $v_{\rm rel}$ is the relative velocity between the hazardous object and the impactor.

\begin{figure}
\includegraphics[width=\linewidth]{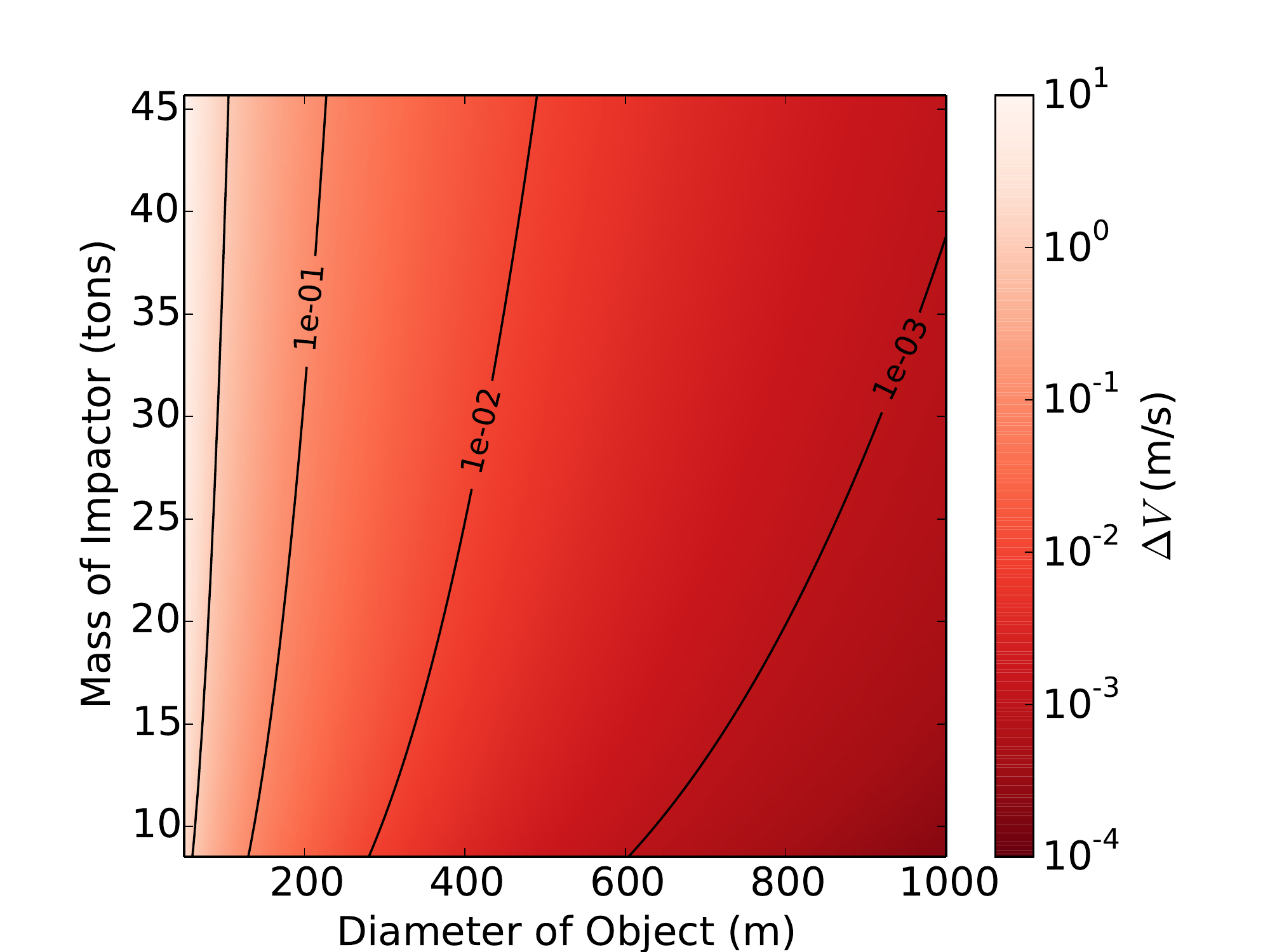}
\caption{\label{fig:deltavkinetic} Predicted $\Delta v$ imparted onto a target object by a kinetic impactor vs. target object size and kinetic impactor mass, assuming a relative velocity of 20 km/s at impact and a $\beta$ parameter of 2. Black solid lines trace contours in $\Delta v$. }
\end{figure}

Figure \ref{fig:deltavkinetic} shows the predicted $\Delta v_{\rm kin}$ values for an impactor with a relative velocity of 20 km/s and our conservative $\beta$ value of $\beta=2$, as a function of target object size and impactor mass. The impactor mass plotted represents the sum of the payload, the dry mass of the payload capsule, and the dry mass of the final booster. In the case of a kinetic impactor in this scenario, the expected $\Delta v_{\rm kin}$ values range from $\sim0.1$ mm/s to $\sim10$ m/s.

\subsubsection{Gravity Tractor}
\label{sec:gravity}

The gravity tractor technique relies on modifying the trajectory of an incoming hazardous object by using the gravitational force exerted by a spacecraft on the object. The spacecraft hovers near the target object and slowly exerts a gravitational ``tug'' in the desired direction with thrusters. In contrast to the previous two methods, the gravity tractor is a slow-push deflection technology (or more accurately, a ``slow-pull'' technology) and requires a substantial lead time to be effective. This method is also the most fuel-expensive of the three we consider, as the spacecraft must not only rendezvous with the object but also match its velocity, and then continue to operate until a successful deflection is achieved. The major advantages of this technique are that it allows for finer control of the target object's trajectory, and is largely insensitive to the internal structure, surface properties, and rotation state of the object \cite{Lu2005}, although the rotation state of an elongated object with a non-uniform gravity field could affect fuel requirements for station-keeping.

The acceleration $a_{\rm trac}$ imparted on a target of radius $r$ by a spacecraft of mass $m_{\rm payload}$ at a distance $d=fr_{\rm obj}$ from the target's surface is
\begin{equation} \label{eqn:gtractor}
a_{\rm trac} = \frac{Gm_{\rm payload}}{(fr_{\rm obj})^2},
\end{equation}
where $G$ is the gravitational constant. For this simple approximation, we neglect the loss of propellant mass as the tractor continues to operate. Note that this approximation means that our simulation will overestimate the effectiveness of the gravity tractor, as the tractor is only effective as long as it can maintain its position relative to the object, and the loss of propellant mass during operation will decrease the acceleration imparted on the object over time.

Equation \ref{eqn:gtractor} suggests that in the optimal case, a spacecraft would be positioned as close to the target as possible ($f<<1$). However, due to potential interactions between the propellant expelled by the spacecraft and the target object's surface, the spacecraft's thrusters must be pointed away from the target. This in turn means that at least two thrusters must be used. Furthermore, in order for the spacecraft to maintain a constant separation $d$ from the target object, the on-board propulsion system must be capable of achieving an effective thrust, $T_{\rm eff}$, greater than the gravitational force imparted on the spacecraft by the target. In other words,
\begin{equation} \label{eqn:Tefflim}
T_{\rm eff} \geq \frac{G m_{\rm payload} m_{\rm obj}}{(fr_{\rm obj})^2}.
\end{equation}
Here we use $T_{\rm eff}$ to indicate the projection of the nominal thrust $T$ onto the radial vector $\vec{r}$ pointing from the spacecraft to the target. For a pair of thrusters, each with plume angle $\phi$, 
\begin{equation} \label{eqn:Teff}
T_{\rm eff} = 2T \cos\{\arcsin(1/f) + \phi\}.
\end{equation}

As mentioned above, the magnitude of the effective thrust is limited due to the necessity to avoid propellant-target interactions. Thus, the spacecraft thrust $T$ presents the limiting factor for how close the spacecraft can get to the target object's surface. Specifically, since the target's acceleration is maximized for minimum $f$, Equations \ref{eqn:Tefflim} and \ref{eqn:Teff} imply that maximum acceleration can be achieved at a minimum $f=f_{\rm min}$, defined such that
\begin{equation}
\frac{2\pi}{3} \frac{G r_{\rm obj} \rho m_{\rm payload}}{f^2_{\rm min} \cos\{\arcsin(1/f)+\phi\}}-T = 0.
\end{equation}
Then the maximum acceleration that the gravity tractor can apply to the target object is 
\begin{equation}
a_{\rm trac} = \frac{G m_{\rm payload}}{(f_{\rm min}r_{\rm obj})^2}.
\end{equation}

\begin{figure}
\includegraphics[width=\linewidth]{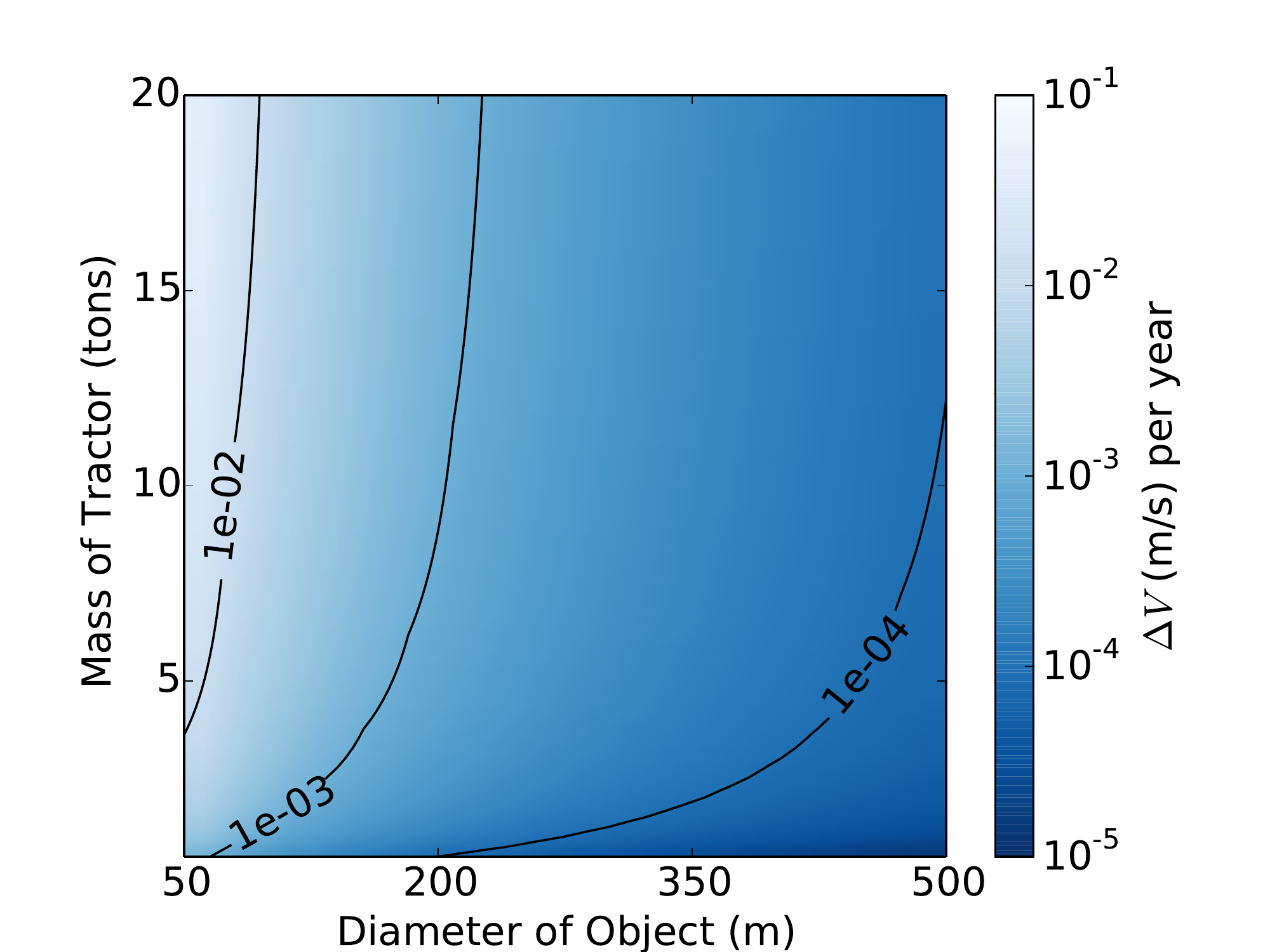}
\caption{\label{fig:deltavtractor} Predicted $\Delta v$ imparted on a target object per year by a gravity tractor with thrust $T=0.8$ N. Black solid lines trace contours in $\Delta v$ per year.}
\end{figure}

Figure \ref{fig:deltavtractor} shows the predicted values for the $\Delta v$ per year imparted by a gravity tractor with a thrust of 0.8 N, as a function of target object size and the mass of the tractor. For this gravity tractor, expected values for $\Delta v$ per year range from $\sim0.01$ mm/s per year to $\sim0.1$ m/s per year.

\subsection{Deflection Simulations}
\label{sec:deflectionsim}

After the backwards integration and the travel time calculation are complete, we then integrate the system forward in time from $t=-15$ yr until time $t=+15$ yr. For every simulated hazardous object, we perform three simulations, each with an attempted deflection by one of our considered technologies. To simulate an attempted deflection, we calculate the maximum possible $\Delta v$ that the  technology can apply to the given object using the equation presented in Section \ref{sec:technologies}. We then add this $\Delta v$ to the object's instantaneous velocity at time $t = t_{\rm lead}$. At each integrator timestep, we check for a collision between the object and the Earth by comparing the object's instantaneous location, $\vec{r}_{\rm obj}$, with the Earth's, $\vec{r}_{\oplus}$. If $\| \vec{r}_{\rm obj}-\vec{r}_{\oplus} \| < R_{\oplus}$, an impact has occurred and we record the simulation as an unsuccessful deflection. If no impact is recorded during the simulation, we record it as a successful deflection. 

The method described above simulates an instantaneous-push deflection, such as a nuclear explosion (Section \ref{sec:nuclear}) or a kinetic impact (Section \ref{sec:kinetic}). We can also simulate a slow-push deflection, such as the deflection by a gravity tractor (Section \ref{sec:gravity}). In slow-push simulations, we apply a change in velocity of $\Delta v/{\rm yr}$ at every integrator timestep $\Delta t$ for which $-t_{\rm lead}<\Delta t<0$. This change in velocity is also applied in the direction of the object's instantaneous velocity vector. We test $\Delta v$/yr values between $10^{-6}$ and $10^2$ m/s per year. 

\begin{figure}
\includegraphics[width=\linewidth]{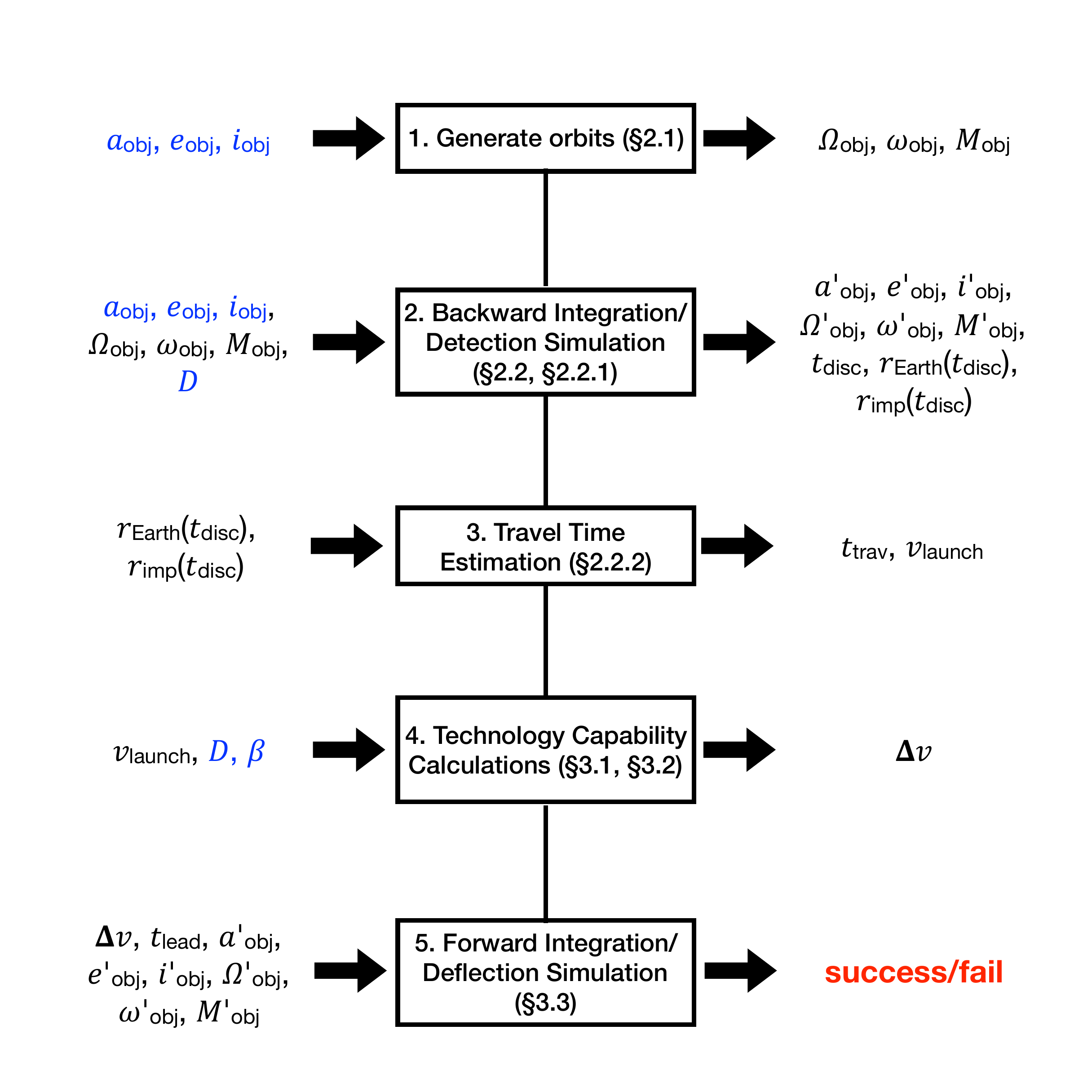}
\caption{\label{fig:diagram2} A diagram summarizing the steps of the Deflector Selector deflection simulation, which produces the training data used in the machine learning algorithm (Section \ref{sec:machinelearning}). The inputs and outputs of each step of the simulation are listed. The inputs highlighted in blue act as the inputs to the machine learning algorithm, while the final output highlighted in red will indicates the classification result of the algorithm. Each step of the algorithm is described in more detail in the section of this paper indicated by the number in parentheses.}
\end{figure}

Figure \ref{fig:diagram2} summarizes the steps of the deflection simulation, specifying the inputs and outputs of each step. For any given hazardous object, only five inputs are needed, those highlighted in blue in Figure \ref{fig:diagram2} and listed in Table \ref{tab:object}. With these five inputs, the Deflector Selector model can predict whether each deflection technology can successfully deflect the object or not.

\begin{figure}
\includegraphics[width=\linewidth]{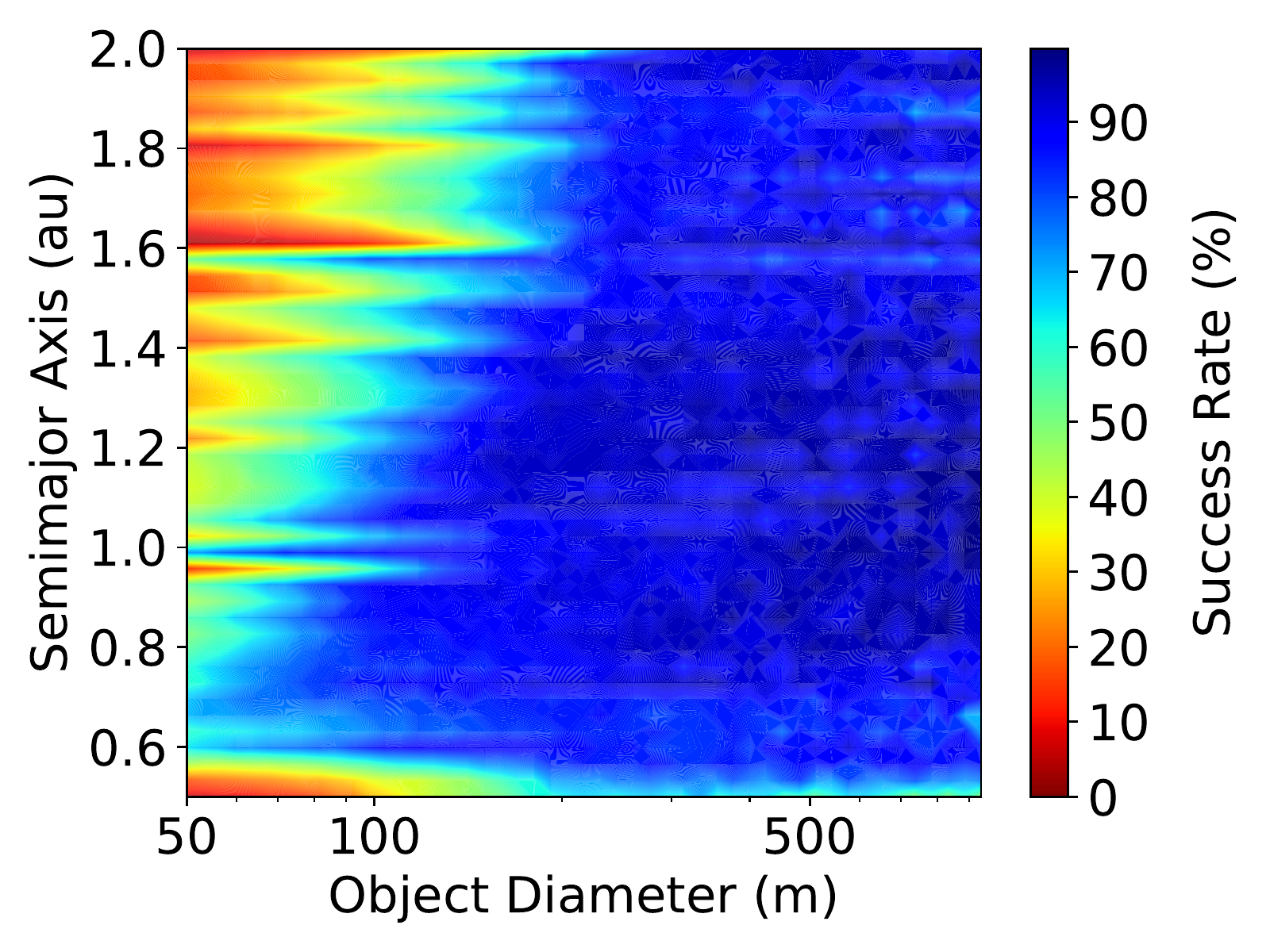}
\caption{\label{fig:success_nuc} Summary of our nuclear explosive deflection simulations. The colors indicate the percentage of successful deflections for a given object diameter and semimajor axis.}
\end{figure}

Our deflection simulations produce a multidimensional data set that can be used to examine the effects of object parameters such as orbit and size on the likelihood that humanity would be capable of deflecting the object before impact. For example, Figure \ref{fig:success_nuc} summarizes the deflection simulations in which a nuclear explosive was attempted. For a given object diameter and semimajor axis, the color of the plot in Figure \ref{fig:success_nuc} indicates the percentage of our simulations in that bin that resulted in successful deflections. Our simulated nuclear explosive can apply a significant $\Delta v$, even to larger objects, so the only constraint on its success is its ability to reach the hazardous object in time. This lead time is a function of the detectability of the object, so smaller objects at larger distances from the Earth tend to have lower success rates. 

\begin{figure}
\includegraphics[width=\linewidth]{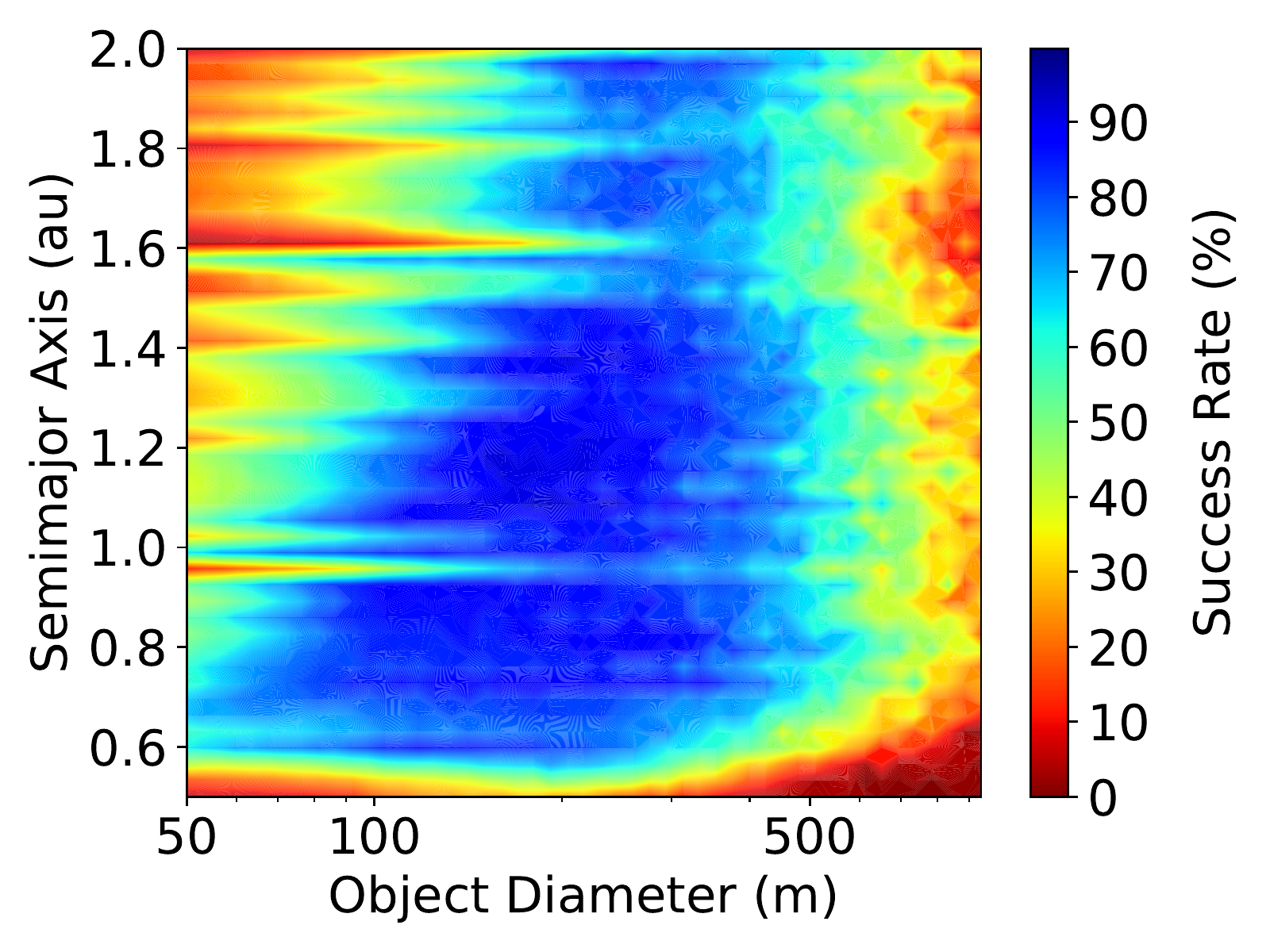}
\caption{\label{fig:success_kin} Summary of our kinetic impactor deflection simulations. The colors indicate the percentage of successful deflections for a given object diameter and semimajor axis.}
\end{figure}

The kinetic impactor success rate, shown in Figure \ref{fig:success_kin}, decreases for both the largest and the smallest objects. For smaller objects, the success rate is lower for objects with semimajor axes farther from the Earth's. The success rate then decreases again for the largest objects. The energy that our simulated kinetic impactor can impart on a hazardous object is smaller than that of a nuclear explosive, so it cannot apply a large enough $\Delta v$ to the largest objects in our sample to successfully deflect them.

\begin{figure}
\includegraphics[width=\linewidth]{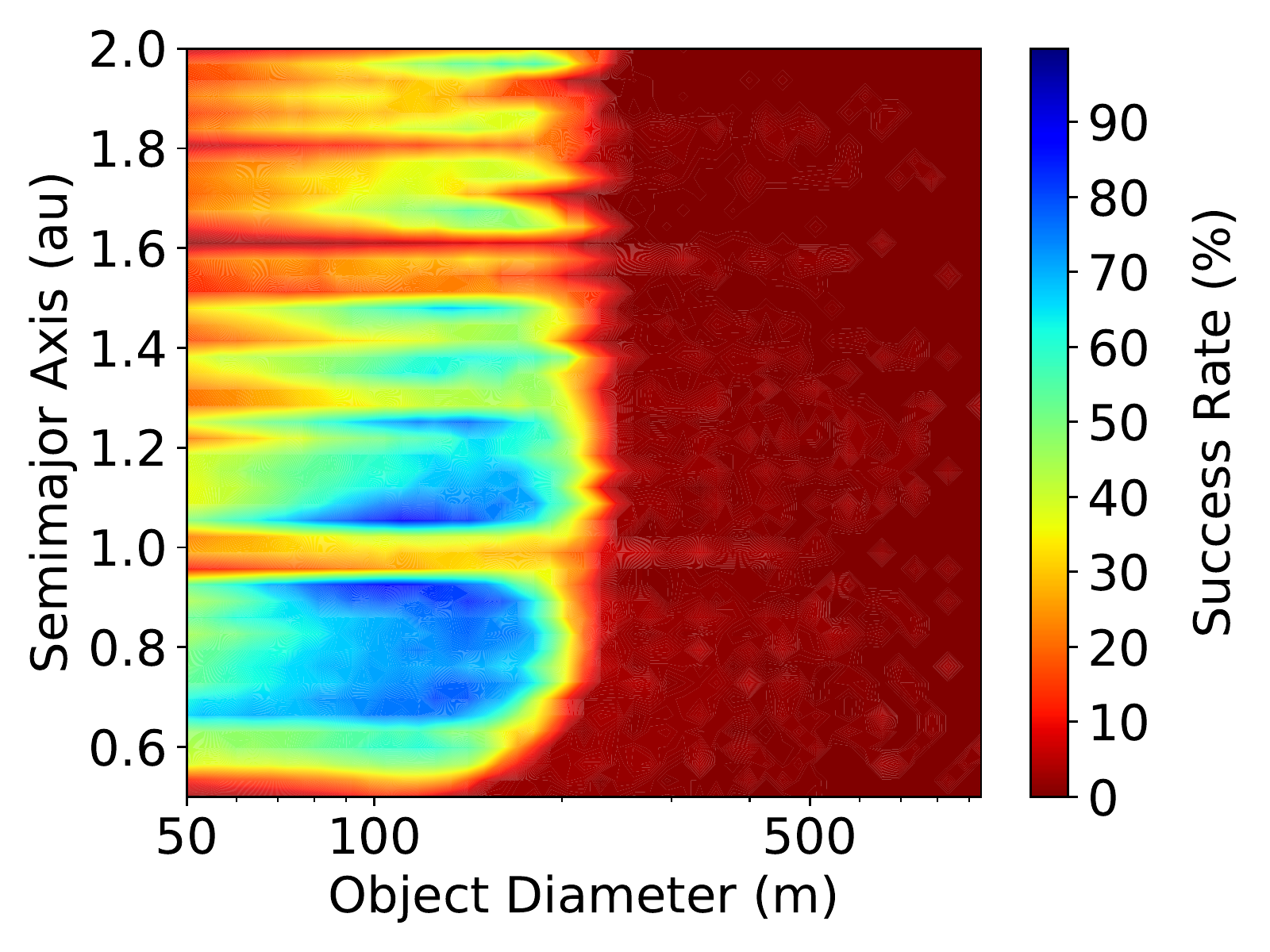}
\caption{\label{fig:success_tra} Summary of our gravity tractor deflection simulations. The colors indicate the percentage of successful deflections for a given object diameter and semimajor axis.}
\end{figure}

The situation is even worse for the gravity tractor, whose success rate we show in Figure \ref{fig:success_tra}. Our simulated gravity tractor can rarely apply a large enough $\Delta v$ to deflect objects larger than $\sim200$ m in diameter.

While Figures \ref{fig:success_nuc}-\ref{fig:success_tra} illustrate the importance of object size and semimajor axis on the probability of a successful deflection, the effects of the object's other parameters (e.g., $\beta$ parameter, eccentricity, inclination) are hidden. To better study the relationships between these parameters and the capabilities of the deflection technologies, we analyzed our results using a machine learning algorithm, described in the next section. 

\section{Machine Learning Algorithm}
\label{sec:machinelearning}

With the exception of object diameter $D$, the hazardous object parameters described in Section \ref{sec:orbital} have uniform distributions and do not represent a realistic distribution of likely Earth-impacting objects. To predict which deflection technologies would be most effective against a realistic simulated population of hazardous objects, the simulation pipeline described in Sections \ref{sec:orbital} and \ref{sec:deflection} can be repeated for different object parameter distributions. However, these calculations are computationally expensive, constraining the user's ability to test more than a few hazardous object populations in a reasonable amount of time. Instead, we used the results of our simulations, summarized in Section \ref{sec:deflectionsim}, to train a machine learning algorithm. The algorithm takes as its input five object features $(a_{\rm obj},e_{\rm obj},i_{\rm obj}, D,$ and $\beta)$ and predicts which combination of technologies are capable of deflecting the object. A user can then feed an object population into the trained algorithm and produce a histogram predicting the effectiveness of each technology against the population. While the training process can be computationally intensive, the trained algorithm can return a result much more quickly than the orbital simulations described above, allowing for rapid comparisons of various simulated object populations.

\subsection{Training Data}
\label{sec:trainingdata}

The result of the deflection simulations described in Section \ref{sec:deflectionsim} is a dataset containing the results of $N_{\rm techs}$ attempted deflections for each of $N_{\rm obj}$ hazardous objects. This dataset contains $N_{\rm obj}$ points, each with the object's $(a_{\rm obj},e_{\rm obj},i_{\rm obj})$, $D$, and $\beta$ parameters, and $N_{\rm techs}=3$ binary flags indicating whether each technology produced a successful deflection. The training data set we used to train the algorithm presented in this work contained $N_{\rm obj}=6$ million samples, representing the results of 18 million simulations. These data took up approximately 136 MB of storage space when saved in a binary file.

\subsection{Decision Trees}

For the machine learning step of our pipeline, we chose to implement a decision tree algorithm. Decision tree methods are effective for classification problems, and have the added advantage of producing a trained algorithm that can be easily interpreted and used to examine the relative importance of the various input features. As it learns, the decision tree algorithm partitions the feature space of the training samples into smaller and smaller regions. For each region of the feature space at each level of the tree, the algorithm must decide how to partition the region to maximize the efficiency of the tree, so it chooses a partition of a certain feature such that the impurity of the training samples in each of the two resulting subregions is minimized. 

We trained and implemented our decision trees using the scikit-learn package of Python \cite{Pedregosa2011}, which uses an optimized version of the Classification and Regression Trees (CART) algorithm \cite{Breiman1984} to build a tree. While training, this decision tree algorithm measures the impurity of each node using the Gini impurity, which is given by 
\begin{equation} \sum\limits_{i=0}^K p_i (1-p_i), \end{equation}
where $K$ is the number of possible classes (in our case, $K=2$) and $p_i$ represents the fraction of samples in the given subset that belong to class $i$. The minimum possible value for the Gini impurity, zero, occurs when every sample in the subset belongs to the same class. 

\begin{figure}
\includegraphics[width=\linewidth]{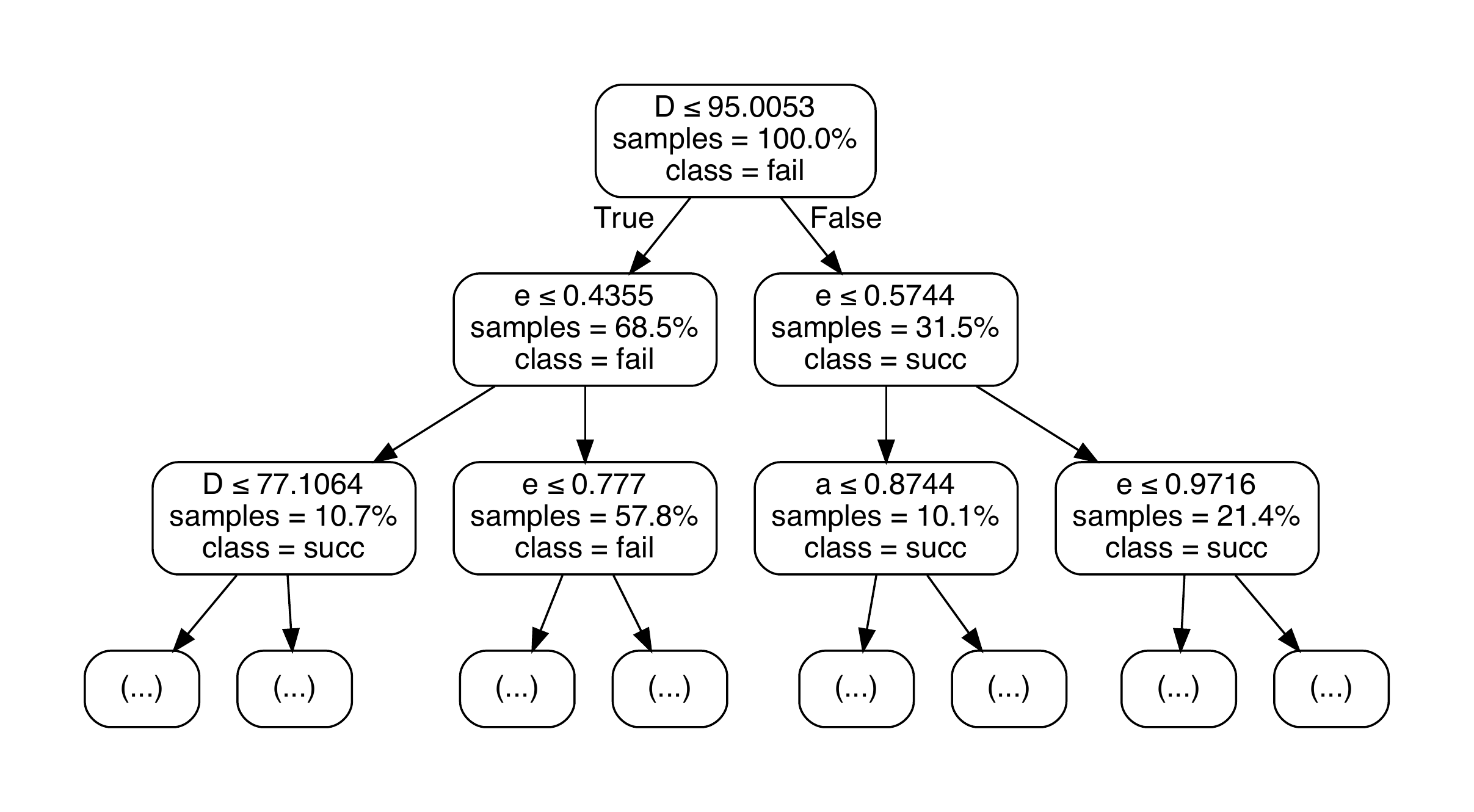}
\caption{\label{fig:treepicture} A graphical representation of the first three layers of an example decision tree trained on the nuclear explosive deflection data. ``Samples'' indicates the percentage of the training samples represented by the given node. ``Class'' indicates the technology classification of the largest fraction of samples in the node. } 
\end{figure}

One benefit of using a decision tree for our machine-learning algorithm is that the trained tree can be visualized and understood by humans, rather than acting as a black box (as in the case of certain other machine-learning algorithms, e.g., neural networks). As an example, in Figure \ref{fig:treepicture}, we illustrate the first few levels of a single decision tree trained on our data for nuclear explosive deflections. In this diagram, each node represents a split by the algorithm according to one of the object parameters. Objects that satisfy the inequality in a given node proceed to the left child node, while objects that do not satisfy the equality move the the right child node. While only the first three layers of the tree are shown here, the tree terminates when each of the nodes has an impurity of zero.  

A single decision tree tends not to be very accurate. In particular, very deep trees can suffer from overfitting. To improve the accuracy of our machine learning algorithm, we implemented an ensemble method knows as a random forest. As it trains on the data, a random forest algorithm builds $N_{\rm trees}>1$ trees. To create a given tree, the algorithm randomly selects a subset of the training data, and when partitioning the data at a node, it chooses the best feature to make the split from a random subset of all the features. After the forest is trained, it classifies a new object but feeding the object's feature through all of its trees and then averaging the result (the various trees in the forest are said to ``vote'' on the classification). The resulting algorithm is a more accurate model than a single decision tree.

We trained three separate random forests on the data, one for each technology. Each forest was constructed with $N_{\rm trees}=100$ trees, each with a maximum depth of 50 levels. We used 80\% of the training data to train each random forest and the remaining 20\% for test their accuracy. To measure the accuracy, we input this test sample into the trained machine learning algorithm, then compared the algorithm's predicted classifications with the known results for the sample from our deflection simulations. Our collection of three random forests had an accuracy of $\sim93$\%. We also experimented with training a single random forest using multiple classification to predict the deflection success of each technology, represented by a single integer, but found that this method tended have an accuracy of only $\sim$80\%. 

The difference in accuracy between a single, muticlass decision tree and three binary decision trees is likely due to the nature of our training data, which is very unbalanced for a multiclass tree. For example, there are no objects in our sample for which the gravity tractor is effective but the nuclear explosive is not. Thus, there are no training data representing this result. However, from the perspective of three binary decision trees, the various classifications are well-represented; there are a large number of samples for which the gravity tractor is successful, and for which the gravity tractor is unsuccessful. Our training data is better balanced, and therefore better suited, for training three separate trees than a single, multiclass tree.

We are interested not only in determining which technologies can successfully deflect a given hazardous object, but also in measuring which object features are most important in making this determination. Some characteristics of a hazardous object can be measured more easily, quickly, and precisely than others, but they may not have as much of an effect on the likelihood that a given technology can successfully deflect the object. For example, if we determined that the decision of which technology could deflect a hazardous object is extremely sensitive to the object's $\beta$ parameter, this would suggest that we need to develop better techniques for measuring the $\beta$ parameter of NEOs. The decision tree algorithm provides us with a tool for understanding the relative importance of each object feature: the Gini importance. As described above, every time the algorithm partitions the training samples based on some object feature, the Gini impurity of the resulting subsets is lower than the initial set. The Gini importance of an object feature is calculated as the normalized total reduction of Gini impurity due to that feature over the entire tree \cite{Breiman2001}. For an ensemble of trees like a random forest, we can calculate the average Gini importance of each feature over the entire set of trees.

\begin{figure}
\includegraphics[scale=0.75]{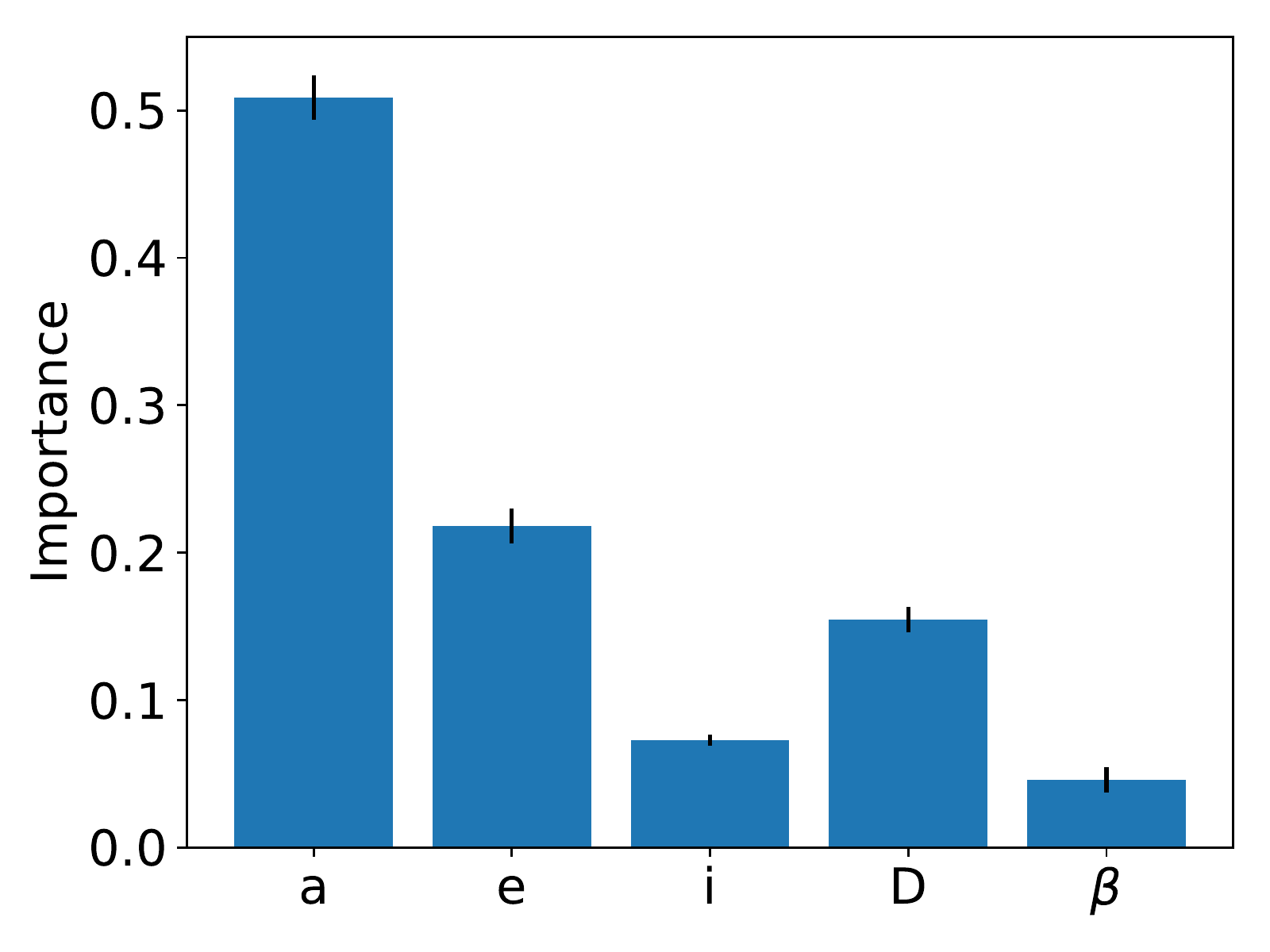}
\centering
\caption{\label{fig:importances} The average Gini importances of each of the five inputs of our machine learning algorithm, with error bars indicating the standard deviation between the three random forests. The object's semimajor axis is the most important determiner of whether each technology can deflect the object.} 
\end{figure}

We plot the Gini importance of each of our five object features in Figure \ref{fig:importances}. An object's semimajor axis is the most important determiner of the technologies that can successfully deflect the object, most likely due to the dependence of the discovery time (and thus the lead time) on semimajor axis. The object's eccentricity and diameter are the next most important features, due to their strong effects on the discovery time, followed by the inclination and $\beta$ parameter, which have little to no effect on the discovery time, and thus the lead time, of the object.

\subsection{Results}
\label{sec:results}

Having trained an algorithm to predict, based on the characteristic of a hazardous object, which technologies can successfully deflect the object, we can now apply the algorithm to a simulated population of objects to predict which technologies would be most useful. In this section we test three simulated object populations and compare the results.

\subsubsection{Realistic Impactor Population}
\label{sec:veres}

We first examine a realistic population of hazardous Near Earth Asteroids (NEAs). We use the simulated impactor population produced by Vere\v{s} et al. \cite{Veres2009}, who produced $10^5$ simulated Earth-impacting orbits from the NEO population model of Bottke et al. \cite{Bottke2000,Bottke2002} using the technique of Chesley \& Spahr \cite{Chesley2004}. We use the Bottke et al. size distribution described in Section \ref{sec:orbitpopulation}. The $\beta$ values of the NEO population is less well-constrained. We use a Gaussian distribution of standard deviation 0.5 around a reasonable value of $\beta=2$.

\begin{figure}
\includegraphics[width=\linewidth]{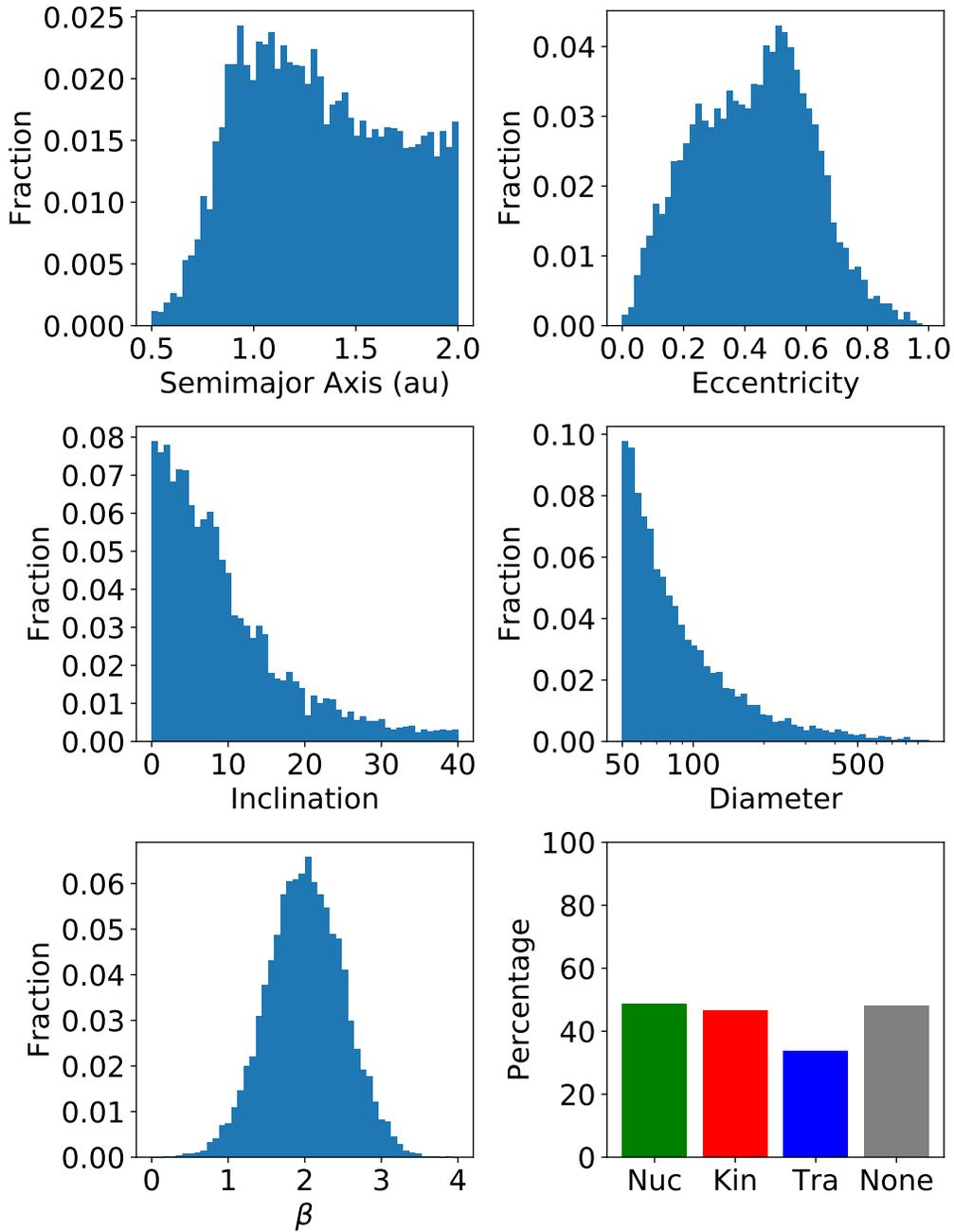}
\caption{\label{fig:techs_veres} Distributions of $a$, $e$, $i$, $D$, and $\beta$ for our simulated Vere\v{s} population of hazardous objects. Lower right: Percentage of the Vere\v{s} population that can be successfully deflected by each technology. The rightmost column indicates the percentage of the population that cannot be successfully deflected by any of the three technologies.} 
\end{figure}

Figure \ref{fig:techs_veres} shows the distributions of the orbital elements, diameter, and $\beta$ parameter for our simulated NEA population, and a plot comparing the success rate of each technology for this population of objects. A nuclear explosive can successfully deflect 49\% of the objects in this population, while 49\% cannot be deflected by any of the three technologies, either because the deflection technology cannot reach the object before impact or because the $\Delta v$ applied by each technology does not change the object's trajectory enough to avoid an impact. The kinetic impactor has a slightly lower success rate than the nuclear explosive (47\%), while the gravity tractor is the least effective technology, with a 33\% success rate. The comparable success rates of the nuclear explosive and the kinetic impactor indicates that the lead time of roughly half of the objects is sufficient for both of these deflection technologies to be effective. The largest of these objects, in the tail of the object size distribution, can be deflected by the large $\Delta v$ of the nuclear explosive but not the smaller $\Delta v$ of the kinetic impactor, despite a large lead time. See Figures \ref{fig:success_nuc} and \ref{fig:success_kin} for a comparison of the capabilities of these two technologies vs. object size.

\subsubsection{Small Comet-Like Objects}
\label{sec:comets}

NEAs are not the only Solar System bodies that pose a threat to the Earth. Comets inbound from the outer Solar System on high-eccentricity orbits can also cross the Earth's orbit and collide with the surface. Such collisions are much less frequent, but can be harder to predict, as these objects will likely not be detected until their final orbit before impact. Additionally, heating of the surface of a comet by a deflection technology could cause off-gassing, which would alter the orbit of the comet in ways that are difficult to predict. For now, we neglect this effect.

\begin{figure}
\includegraphics[width=\linewidth]{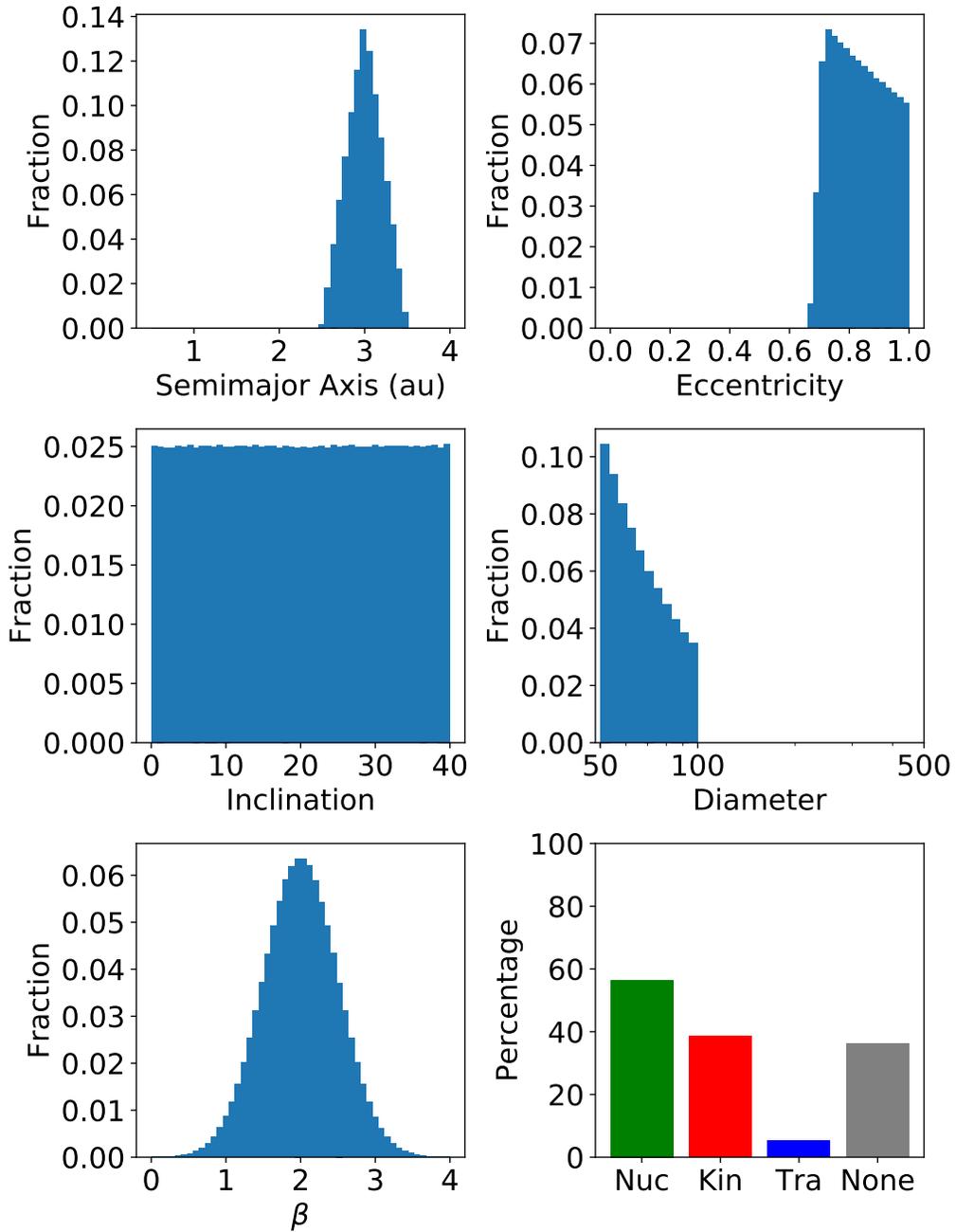}
\caption{\label{fig:techs_comet} Distributions of $a$, $e$, $i$, $D$, and $\beta$ for our simulated small comet hazardous object population. Lower right: Percentage of our simulated comets that can be successfully deflected by each technology. The rightmost column indicates the percentage of the population that cannot be successfully deflected by any of the three technologies.} 
\end{figure}

The characteristics of the long-period comet population in the Solar System is poorly constrained, and their semi-major axes are far larger than the range we used to train our machine learning algorithm. So for now we simulate a population of Jupiter-family-like comets with apocenter distances uniformly distributed between 5 and 6 au and perihelion distances constrained to be less than 1 au, to create Earth-crossing orbits. This forces the eccentricities of the objects to be between 0.83 and 0.87. We use a uniform distribution of inclinations between 0 and $40^{\circ}$, and focus on the smaller hazardous objects by using a Bottke size distribution between 50 and 100 m in diameter. We use the same Gaussian distribution of $\beta$ values around $\beta=2$, as described in Section \ref{sec:veres}. These feature distributions, and the technology success rates, are summarized in Figure \ref{fig:techs_comet}. 

Nuclear explosives are moderately effective on this population, with a success rate of 56\%, likely due to the small sizes of the objects. However, kinetic impactors and gravity tractors are much less effective, with success rates of 39\% and 5\%, respectively. This is likely due to the orbits of the objects, which makes them more difficult to detect, decreasing their lead time. The $\Delta v$ applied by a kinetic impactor is not large enough in all cases to deflect the object's trajectory in the available lead time. The gravity tractor's total applied $\Delta v$ increases with lead time, so for shorter lead times, it may not be able to deflect the object at all.

\subsubsection{Rubble Piles}
\label{sec:rubble}

\begin{figure}
\includegraphics[width=\linewidth]{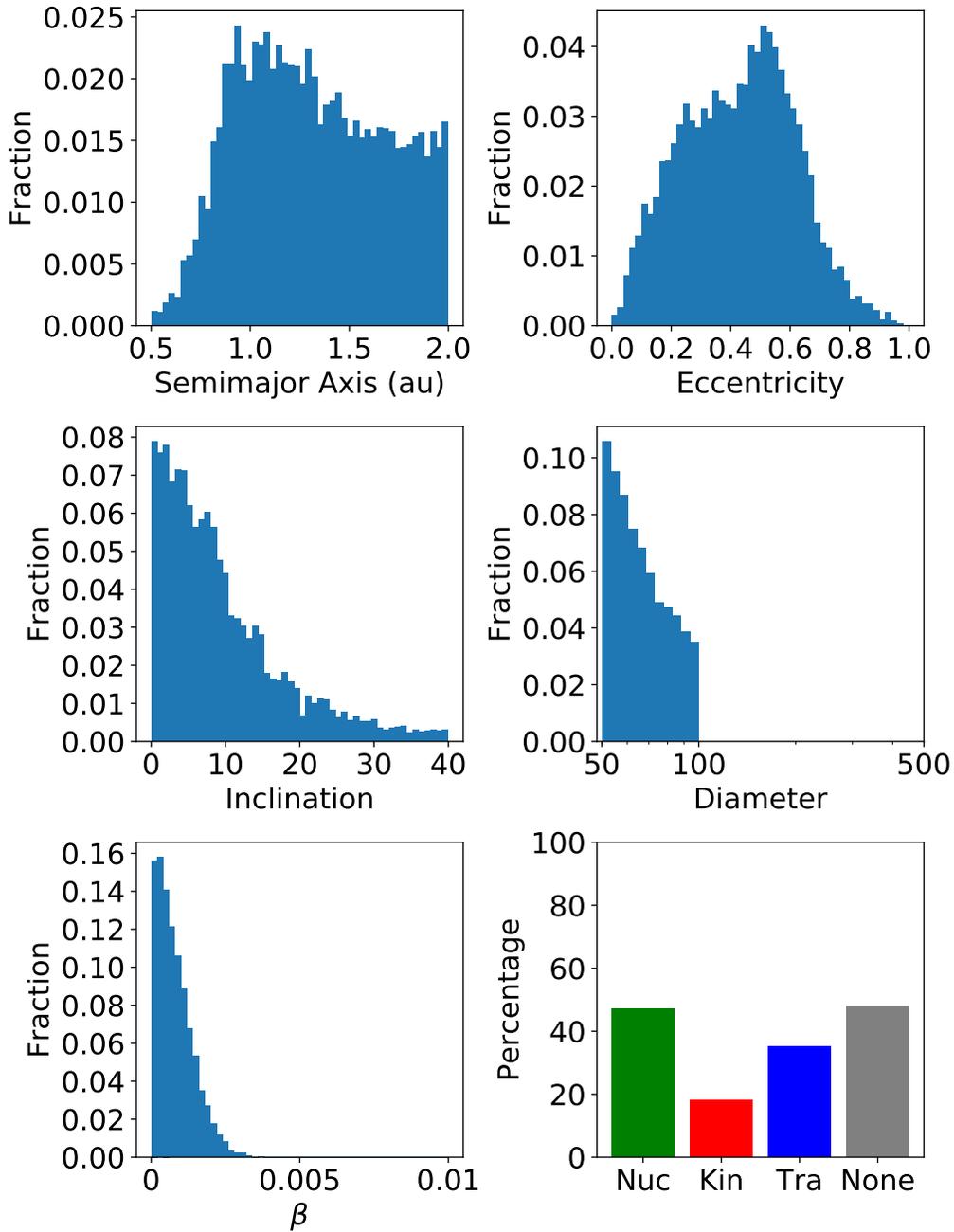}
\caption{\label{fig:techs_rubble} Distributions of $a$, $e$, $i$, $D$, and $\beta$ for our simulated small rubble pile hazardous object population. Lower right: Percentage of our simulated population of rubble piles that can be successfully deflected by each technology. The rightmost column indicates the percentage of the population that cannot be successfully deflected by any of the three technologies.} 
\end{figure}

Finally, we explored the effect of the $\beta$ parameter on the success rate of the deflection technologies. The $\Delta v$ applied by a nuclear explosive or kinetic impactor depends on the value of $\beta$ (see Equations \ref{eqn:nuclear} and \ref{eqn:kinetic}), although this dependence is much stronger for the kinetic impactor, as only the debris term of the nuclear explosive's $\Delta v$ depends on $\beta$. We tested a population of small hazardous objects with orbits drawn from the Vere\v{s} simulated hazardous object population, diameters following a Bottke distribution between 50 and 100 m, but $\beta$ values distributed in the positive half of a Gaussian distribution with a mean of zero and a standard deviation of 0.001. These $\beta$ values close to zero represent objects with very low internal strengths: rubble piles held together loosely, rather than solid bodies. The parameter distributions of these small near-Earth rubble piles and the resulting technology success rates are shown in Figure \ref{fig:techs_rubble}.

Our results indicate that the success rate of the nuclear explosive does not decrease significantly when the NEAs are replaced with rubble piles. However, the effectiveness of the kinetic impactor drops to 18\% due to the dependence of $\Delta v$ on $\beta$. The gravity tractor is slightly more effective on this population (35\%) due to the smaller size of the objects.

\section{Discussion and Conclusions}
\label{sec:conclusions}

We have presented our hazardous object deflection model, the Deflector Selector, which consists of a machine learning algorithm trained on data produced by orbital integrations simulating deflection attempts. In this work, we tested the effectiveness of three deflection technologies on the simulated hazardous object populations. Unsurprisingly, our results indicate that nuclear explosives are the most effective deflection technology regardless of the object population, although the effectiveness of nuclear explosives does vary somewhat with the object's orbit and diameter. However, nuclear explosives may not be the most practical approach to deflecting an incoming hazardous object, due to the threat posed by a catastrophic launch failure. The nuclear fallout from the accidental explosion of a launch vehicle carrying a nuclear deflector to space could result in significant loss of life and damage to property. A kinetic impactor can be nearly as effective as nuclear explosives on certain populations of hazardous objects, but its effectiveness depends on the internal strength of the objects, represented in our model by the $\beta$ parameter. The gravity tractor is insensitive to $\beta$, but its effectiveness is highly sensitive to the lead time before impact, which is a function of both the object's orbit and its size.

While the orbital simulations described in Sections \ref{sec:orbital} and \ref{sec:deflection} can be used to directly study the effectiveness of various deflection technologies, they are computationally expensive. Using a computer cluster with 100 cores, we required $\sim40$ hours of wall time to complete the simulations described in this work. This computational cost limits the ability of the model to test a variety of object populations to compare the effectiveness of technologies. Once trained, our decision tree algorithm can perform the same classifications in seconds.

\subsection{Future Work}
\label{sec:future}

In this work, we have described the framework the Deflector Selector pipeline and examined the effectiveness of three of the more plausible deflection technologies. Future modifications to the pipeline will improve its accuracy and its applicability to a wider range of objects and technologies.

For example, we plan to increase the parameter space for the hazardous objects in future versions of the model. This will require additional orbital simulations to test the deflection of objects with a larger range of sizes and orbits. To study the threat of short-period comets from the Kuiper belt or long-period comets from the Oort cloud, we will need to simulate the deflection of objects with much larger semimajor axes. Comets can also be much larger than asteroids, so we will also need to test object diameters greater than 1 km.

We also plan to refine the approximations made in the simulation. For example, we have assumed in this work that a potentially hazardous object's orbit can be characterized upon its discovery, such that it is instantly identified as a threat. In reality, multiply apparitions are required to characterize an object's orbit, and the time required to determine whether an object is a threat depends strongly on its size \cite{Chodas2008}. We will add these considerations to the lead time simulations in future versions of the model. We will also experiment with different values for the albedo of an object. In this work, we used a single albedo value of 0.25 to relate the magnitude of an object with its diameter. Recent observations with NEOWISE indicate a bimodal distribution of albedos in the Near Earth Asteroids (NEAs), with $\sim25\%$ of the NEAs making up a very dark population with a peak albedo of 0.03, and $\sim75\%$ of NEAs making up a moderately dark population with a peak albedo of 0.168 \cite{Wright2016}. The higher albedo used in our model means that an object with a given diameter will have a lower apparent magnitude, making them easier to detect. In the future, we will explore the effects of using lower albedos, and using an orbit-dependent albedo distribution.

The current version of the model only attempts deflections in which the change in velocity $\Delta v$ is positive, i.e., applied along the direction of the hazardous object's instantaneous velocity. Future versions of the model will explore the effects of other types of $\Delta v$ deflections, in particular, a negative $\Delta v$ applied opposite to the object's direction of motion. This will allow a more accurate simulation of deflection technologies that depend strongly on the relative velocities of the deflector and the hazardous object, such as the kinetic impactor.

In the future, by working with technology researchers and engineers, we will vary the capabilities of the three technologies we have tested here to address specific technology proposals. For example, we can test gravity tractors with more powerful thrusters, or rockets with larger capacities. We also plan to incorporate additional deflection technologies into future versions of the model, including mass drivers, laser ablation, and ion beam shepherding. Some of these technologies will require the addition of other object characteristics into the model. For example, mass drivers are anchored to the surface of an object, extract material, and eject the material into space to apply an acceleration to the object. Both the landing on and the ejection of material from the surface of the object are governed by the rotation period of the object, so this period must be considered as an additional object feature in our model. Other object parameters, such as albedo, will affect the discovery time and thus the lead time for a given object, and will be included in future versions.

Using further simulations to produce more data for training the machine learning algorithm will likely have both positive and negative effects on the accuracy of the algorithm. Higher-resolution coverage of the parameter space may allow the algorithm to identify additional structure in this space, increasing its accuracy. However, expanding the parameter space by increasing the complexity of the model may reduce the algorithm's accuracy unless the new regions of parameter space can be adequately covered by new training data.

\vspace{10pt}

\noindent Funding: This work was supported by the National Aeronautics and Space Administration Frontier Development Lab. The orbital simulations were performed on the Memex High Performance Computing Cluster at the Carnegie Institution for Science.

\vspace{10pt}

\bibliographystyle{elsarticle-num}
\bibliography{./DeflectorSelector}

\begin{thebibliography}{10}
\expandafter\ifx\csname url\endcsname\relax
  \def\url#1{\texttt{#1}}\fi
\expandafter\ifx\csname urlprefix\endcsname\relax\def\urlprefix{URL }\fi
\expandafter\ifx\csname href\endcsname\relax
  \def\href#1#2{#2} \def\path#1{#1}\fi

\bibitem{Schulte2010}
P.~Schulte, L.~Alegret, I.~Arenillas, J.~A. Arz, P.~J. Barton, P.~R. Bown,
  T.~J. Bralower, G.~L. Christeson, P.~Claeys, C.~S. Cockell, G.~S. Collins,
  A.~Deutsch, T.~J. Goldin, K.~Goto, J.~M. Grajales-Nishimura, R.~A.~F. Grieve,
  S.~P.~S. Gulick, K.~R. Johnson, W.~Kiessling, C.~Koeberl, D.~A. Kring, K.~G.
  MacLeod, T.~Matsui, J.~Melosh, A.~Montanari, J.~V. Morgan, C.~R. Neal, D.~J.
  Nichols, R.~D. Norris, E.~Pierazzo, G.~Ravizza, M.~Rebolledo-Vieyra, W.~U.
  Reimold, E.~Robin, T.~Salge, R.~P. Speijer, A.~R. Sweet,
  J.~Urrutia-Fucugauchi, V.~Vajda, M.~T. Whalen, P.~S. Willumsen, {The
  Chicxulub Asteroid Impact and Mass Extinction at the Cretaceous-Paleogene
  Boundary}, Science 327~(5970) (2010) 1214--1218.

\bibitem{Black2013}
P.~Black, L.~Smith-Spark, {Russia starts cleanup after meteor strike} (feb
  2013).

\bibitem{Larson1998}
S.~Larson, J.~Brownlee, C.~Hergenrother, T.~Spahr, {The Catalina Sky Survey for
  NEOs}, in: American Astronomical Society, DPS meeting {\#}30, 1998.

\bibitem{Stokes2000}
G.~H. Stokes, J.~B. Evans, H.~E.~M. Viggh, F.~C. Shelly, E.~C. Pearce, {Lincoln
  Near-Earth Asteroid Program (LINEAR)}, Icarus 148 (2000) 21--28.

\bibitem{Wainscoat2016}
R.~Wainscoat, {The Pan-STARRS search for Near Earth Objects}, in: 2016 IEEE
  Aerospace Conference, 2016, pp. 1--6.

\bibitem{Mainzer2011}
A.~Mainzer, J.~Bauer, T.~Grav, J.~Masiero, R.~M. Cutri, J.~Dailey,
  P.~Eisenhardt, R.~S. McMillan, E.~Wright, R.~Walker, R.~Jedicke, T.~Spahr,
  D.~Tholen, R.~Alles, R.~Beck, H.~Brandenburg, T.~Conrow, T.~Evans, J.~Fowler,
  T.~Jarrett, K.~Marsh, F.~Masci, H.~McCallon, S.~Wheelock, M.~Wittman,
  P.~Wyatt, E.~DeBaun, G.~Elliott, D.~Elsbury, T.~Gautier, S.~Gomillion,
  D.~Leisawitz, C.~Maleszewski, M.~Micheli, A.~Wilkins, {PRELIMINARY RESULTS
  FROM NEOWISE: AN ENHANCEMENT TO THE WIDE-FIELD INFRARED SURVEY EXPLORER FOR
  SOLAR SYSTEM SCIENCE}, The Astrophysical Journal 731~(1) (2011) 53.

\bibitem{Hammerling1995}
P.~Hammerling, J.~L. Remo, {NEO interaction with nuclear radiation}, Acta
  Astronautica 36 (1995) 337--346.

\bibitem{Dachwald2007}
B.~Dachwald, B.~Wie, {Solar Sail Kinetic Energy Impactor Trajectory
  Optimization for an Asteroid-Deflection Mission}, Journal of Spacecraft and
  Rockets 44~(4) (2007) 755--764.

\bibitem{Vasile2008}
M.~Vasile, C.~Colombo, {Optimal Impact Strategies for Asteroid Deflection},
  Journal of Guidance, Control, and Dynamics 31~(4) (2008) 858--872.

\bibitem{Lu2005}
E.~T. Lu, S.~G. Love, {Gravitational tractor for towing asteroids}, Nature 438
  (2005) 177--178.

\bibitem{Melosh1994}
H.~J. Melosh, I.~V. Nemchinov, Y.~I. Zetzer, {Non-nuclear strategies for
  deflecting comets and asteroids}, University of Arizona Press, 1994.

\bibitem{Lubin2016}
P.~Lubin, G.~B. Hughes, M.~Eskenazi, K.~Kosmo, I.~E. Johansson, J.~Griswold,
  M.~Pryor, H.~O'Neill, P.~Meinhold, J.~Suen, J.~Riley, Q.~Zhang, K.~Walsh,
  C.~Melis, M.~Kangas, C.~Motta, T.~Brashears, {Directed energy missions for
  planetary defense}, Advances in Space Research 58 (2016) 1093--1116.

\bibitem{Bombardelli2013}
C.~Bombardelli, H.~Urrutxua, M.~Merino, J.~Pel{\'{a}}ez, E.~Ahedo, {The ion
  beam shepherd: A new concept for asteroid deflection}, Acta Astronautica 90
  (2013) 98--102.
\newblock \href {http://dx.doi.org/10.1016/j.actaastro.2012.10.019}
  {\path{doi:10.1016/j.actaastro.2012.10.019}}.

\bibitem{AHearn2005}
M.~F. A'Hearn, M.~J.~S. Belton, W.~A. Delamere, J.~Kissel, K.~P. Klaasen, L.~A.
  McFadden, K.~J. Meech, H.~J. Melosh, P.~H. Schultz, J.~Sunshine, P.~C.
  Thomas, J.~Veverka, D.~K. Yeomans, M.~W. Baca, I.~Busko, C.~J. Crockett,
  S.~M. Collins, M.~Desnoyer, C.~A. Eberhardy, C.~M. Ernst, T.~Farnham,
  L.~Feaga, O.~Groussin, D.~Hampton, S.~I. Ipatov, J.~Y. Li, D.~Lindler,
  C.~Lisse, N.~Mastrodemos, W.~M. Owen, J.~E. Richardson, D.~Wellnitz, R.~L.
  White, {Deep Impact: Excavating Comet Tempel 1.}, Science 310~(5746) (2005)
  258--264.

\bibitem{Nations1966}
{United Nations}, {Treaty and Principles Governing the Activities of States in
  the Exoploration and Use of Outer Space, including the Moon and Other
  Celestial Bodies} (1966).

\bibitem{Carusi2002}
A.~Carusi, G.~B. Valsecchi, G.~D'Abramo, A.~Boattini, {Deflecting NEOs in Route
  of Collision with the Earth}, Icarus 159 (2002) 417--422.

\bibitem{SanchezCuartielles2010}
J.~{Sanchez Cuartielles}, {Asteroid Hazard Mitigation: Deflection Models and
  Mission Analysis}, Ph.D. thesis, University of Glasgow (2010).

\bibitem{SanchezCuartielles2007}
J.~P. {Sanchez Cuartielles}, C.~Colombo, M.~Vasile, G.~Radice, {A
  Multi-criteria Assessment of Deflection Methods for Dangerous NEOs}, in: AIP
  Conference Proceedings, 2007, pp. 317--334.

\bibitem{Bottke2000}
W.~F. Bottke, R.~Jedicke, A.~Morbidelli, J.-M. Petit, B.~Gladman,
  {Understanding the Distribution of Near-Earth Asteroids}, Science 288~(5474)
  (2000) 2190--2194.

\bibitem{Fowler1992}
J.~W. Fowler, J.~R. Chillemi, {The IRAS Minor Planet Survey}, Tech. Rep.
  PL-TR-92-204, Phillips Laboratory, Hanscom AF Base, MA (1992).

\bibitem{Rein2012}
H.~Rein, S.~Liu, {REBOUND: an open-source multi-purpose N -body code for
  collisional dynamics}, Astronomy {\&} Astrophysics 537 (2012) A128.

\bibitem{Galache2015}
J.~L. Galache, C.~L. Beeson, K.~K. McLeod, M.~Elvis, {The need for speed in
  near-earth asteroid characterization}, Planetary and Space Science 111 (2015)
  155--166.

\bibitem{Ahrens1992}
T.~J. Ahrens, A.~w. Harris, {Deflection and fragmentation of near-Earth
  asteroids}, Nature 360 (1992) 429--433.

\bibitem{Hall1997}
C.~D. Hall, I.~M. Ross, {Dynamics and control problems in the deflection of
  near-earth objects}, in: Proceedings of the 1997 AIAA/AAS Astrodynamics
  Conference, Sun Valley, Idaho, 1997, pp. 1--18.

\bibitem{Bate1971}
R.~R. Bate, D.~D. Mueller, J.~E. White, {Fundamentals of Astrodynamics}, Dover
  Publications, Inc., New York, 1971.

\bibitem{Vallado1997}
D.~A. Vallado, {Fundamentals of Astrodynamics and Applications}, McGraw-Hill
  Companies, Inc., New York, 1997.

\bibitem{NewHorizons2007}
{Johns Hopkins University Applied Physics Laboratory}, {New Horizons: The First
  Mission to Pluto and the Kuiper Belt: Exploring Frontier Worlds}, Press Kit.

\bibitem{Pedregosa2011}
F.~Pedregosa, G.~Varoquaux, A.~Gramfort, V.~Michel, B.~Thirion, O.~Grisel,
  M.~Blondel, P.~Prettenhofer, R.~Weiss, V.~Dubourg, J.~Vanderplas, A.~Passos,
  D.~Cournapeau, M.~Brucher, M.~Perrot, {\'{E}}.~Duchesnay, {Scikit-learn:
  Machine Learning in Python}, Journal of Machine Learning Research 12 (2011)
  2825--2830.

\bibitem{Breiman1984}
L.~Breiman, J.~H. Friedman, R.~A. Olshen, C.~J. Stone, {Classification and
  regression trees}, Wadsworth, Belmost, CA, 1984.

\bibitem{Breiman2001}
L.~Breiman, {Random Forests}, Machine Learning 45~(1) (2001) 5--32.

\bibitem{Veres2009}
P.~Vere{\v{s}}, R.~Jedicke, R.~Wainscoat, M.~Granvik, S.~Chesley, S.~Abe,
  L.~Denneau, T.~Grav, {Detection of Earth-impacting asteroids with the next
  generation all-sky surveys}, Icarus 203~(2) (2009) 472--485.

\bibitem{Bottke2002}
W.~F. Bottke, A.~Morbidelli, R.~Jedicke, J.-M. Petit, H.~F. Levison, P.~Michel,
  T.~S. Metcalfe, {Debiased Orbital and Absolute Magnitude Distribution of the
  Near-Earth Objects}, Icarus 156~(2) (2002) 399--433.

\bibitem{Chesley2004}
S.~R. Chesley, T.~B. Spahr, {Earth impactors: orbital characteristics and
  warning times}, in: M.~Belton, T.~H. Morgan, N.~Samarasinha, D.~K. Teomans
  (Eds.), Mitigation of hazardous comets and asteroids, Cambridge University
  Press, Cambridge, UK, 2004, p.~22.

\bibitem{Chodas2008}
P.~W. Chodas, J.~D. Giorgini, {Mitigation of Hazardous Comets and Asteroids},
  in: Aseroids, Comets, Meteors 2008, Baltimore, MD, 2008, p. 8371.

\bibitem{Wright2016}
E.~L. Wright, A.~Mainzer, J.~Masiero, T.~Grav, J.~Bauer, {The Albedo
  Distribution of Near Earth Asteroids}, The Astronomical Journal 152 (2016)
  79.

\end{thebibliography}

\end{document}